\documentclass[twocolumn]{aastex62}

\shorttitle{FAST Spectrograph Archive}
\shortauthors{Mink et al.}

\begin{document}

\newcommand{\kms}{km~s$^{-1}$}

\title{Center for Astrophysics Optical Infrared Science Archive I. FAST Spectrograph}

\correspondingauthor{Jessica Mink}
\email{jmink@cfa.harvard.edu}

\author[0000-0003-3594-1823]{Jessica Mink}
\affil{Smithsonian Astrophysical Observatory,
60 Garden St., Cambridge, MA 02138, USA}

\author[0000-0002-4462-2341]{Warren R. Brown}
\affil{Smithsonian Astrophysical Observatory,
60 Garden St., Cambridge, MA 02138, USA}

\author[0000-0002-7924-3253]{Igor V. Chilingarian}
\affil{Smithsonian Astrophysical Observatory,
60 Garden St., Cambridge, MA 02138, USA}
\affil{Sternberg Astronomical Institute,
M.V. Lomonosov Moscow State University, Universitetsky prospect 13, Moscow, 119234, Russia}

\author[0000-0002-1311-4942]{Daniel Fabricant}
\affil{Smithsonian Astrophysical Observatory,
60 Garden St., Cambridge, MA 02138, USA}

\author[0000-0002-6949-0090]{Michael J. Kurtz}
\affil{Smithsonian Astrophysical Observatory,
60 Garden St., Cambridge, MA 02138, USA}

\author[0000-0002-9194-5071]{Sean Moran}
\affil{Smithsonian Astrophysical Observatory,
60 Garden St., Cambridge, MA 02138, USA}

\author[0000-0001-9214-7437]{Jaehyon Rhee}
\affil{Smithsonian Astrophysical Observatory,
60 Garden St., Cambridge, MA 02138, USA}

\author{Susan Tokarz}
\affil{Smithsonian Astrophysical Observatory,
60 Garden St., Cambridge, MA 02138, USA}

\author{William F. Wyatt}
\affil{Smithsonian Astrophysical Observatory,
60 Garden St., Cambridge, MA 02138, USA}

\begin{abstract}
We announce the public release of 141,531 moderate-dispersion optical spectra of 72,247 objects acquired over the past 25 years with the FAST Spectrograph on the Fred L.\ Whipple Observatory 1.5-meter Tillinghast telescope.  We describe the data acquisition and processing so that scientists can understand the spectra.  We highlight some of the largest FAST survey programs, and make recommendations for use.  The spectra have been placed in a Virtual Observatory accessible archive and are ready for download.
\end{abstract}


\section{Introduction} \label{sec:intro}

The FAst Spectrograph for the Tillenghast Telescope (FAST) \citep{fabricant1998} is the workhorse moderate-dispersion spectrograph on the Fred L.\ Whipple Observatory (FLWO) 1.5m Tillinghast telescope, located on Mt. Hopkins, Arizona.  Astronomers at the Center for Astrophyics $|$ Harvard \& Smithsonian (hereafter CfA) have used FAST to obtain over 160,000 spectra since it began operation in 1994 January.  Built to continue work on the CfA Redshift Survey \citep{geller89}, FAST has taken spectra for the Updated Zwicky Catalog \citep{falco99} and the northern part of the 2MASS Redshift Survey \citep{huchra12}, classified thousands of supernovae \citep{matheson08, blondin12, hicken17}, monitored symbiotic stars \citep{kenyon2016} and active galactic nuclei \citep{trichas12}, and acquired data for hundreds of other scientific programs.

The goal of this paper is to enable users to understand FAST spectra in order to use the reduced data archive.  The Optical Infrared (OIR) Telescope Data Center at the Smithsonian Astrophysical Observatory (SAO) has processed, archived, and distributed over 160,000 raw FAST spectra in the last 25 years.

A Virtual Observatory-accessible archive of reduced FAST spectra is available at the CfA Optical/Infrared Science Archive \url{http://oirsa.cfa.harvard.edu}.  The reduced FAST spectra archive has a number of strengths that include stable spectrograph operations, calibrations, and data processing over the past quarter century.  However, the FAST archive is not a uniform set of survey data, and so it is important to understand the data before using it.

We begin with a brief description of the FAST spectrograph and the observing protocols used to acquire raw data.  We then discuss the data reduction pipeline used to extract one-dimensional spectra, and the cross-correlation analysis used to derive radial velocities.  We present example spectra, and highlight the largest data sets in the FAST archive.  We close by describing how to access the data and recommend data quality flags.  All of the spectra are stored in standard FITS files.  We document the cross-correlation templates used to estimate radial velocities in Appendix B and the full list of the keywords stored in each FITS image header in Appendix C.

\begin{figure}
 \includegraphics[width=3.5in]{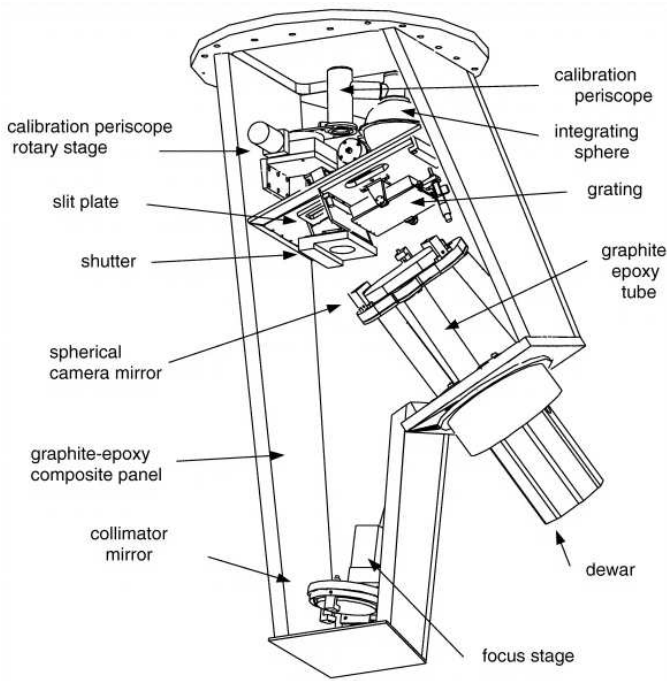}
 \caption{ \label{fig:fast}
	Schematic of the FAST spectrograph.  Light enters from the top, passes through the slit plate, reflects off of the collimator mirror at the bottom, is dispersed by a reflection grating, and then is imaged by a folded Schmidt camera onto a CCD detector.  The CCD dewar extends to the bottom right. Adapted from \citet{fabricant1998}.}
\end{figure}

\section{The Spectrograph}

FAST is a long-slit optical spectrograph described in detail by \citet{fabricant1998}.  Figure \ref{fig:fast} presents a schematic of the instrument.  Light from the telescope enters from above and is focused onto the slit plate, where a slit-viewing camera enables the observer to align an astronomical target onto the slit.  Light that passes through the slit is reflected off a collimator mirror, is dispersed by a reflection grating, and then is imaged by a folded Schmidt camera onto a CCD detector.  An integrating sphere with an incandescent and a helium-neon-argon lamp can be used to illuminate the slit for flat-field and wavelength calibrations, respectively.  

\subsection{Configurations}

The spectrograph is used in different configurations defined by the choice of grating, the grating tilt angle, and slit size.  A fourth spectrograph setting, detector binning, defines the sampling in the spatial direction.  These settings are recorded in the FITS headers as DISPERSE, TILTPOS, APERTURE, and BIN, respectively.

\begin{deluxetable}{cccccr}   
\tablecolumns{6}
\tablewidth{0pt}
\tablecaption{Top Four FAST Configurations\tablenotemark{a}\label{tab:fastconfig}}
\tablehead{
     \colhead{Grating} & \colhead{Tilt} & \colhead{Slit} & \colhead{Coverage} & \colhead{Res.} & \colhead{\%} \\
     (l mm$^{-1}$) & (mil) & ($^{\prime\prime}$) & (\AA) & (\AA) & }
\startdata
	300 & 590 & 3.0 & 3475---7415 & 7.2 & 81\% \\
    600 & 445 & 1.5 & 3540---5530 & 1.8 & 13\% \\
    600 & 752 & 2.0 & 5570---7570 & 2.3 & 2\% \\
   1200 & 850 & 2.0 & 6200---7200 & 1.1 & 1\% 
\enddata
\tablenotetext{a}{Ordered by percentage of spectra obtained from 2006 June through 2019 December.}
\end{deluxetable}

\begin{itemize}
\item FAST has three reflection gratings: a 300 line mm$^{-1}$ and a 600 line mm$^{-1}$ grating blazed at 4750 \AA, plus a 1200 line mm$^{-1}$ grating blazed at 5700 \AA.  Blocking filters are available but in practice never used, with the exception of settings centered around H$_\alpha$ using the 1200 line mm$^{-1}$ grating.  Second-order light is thus present above $\sim$6500~\AA\ in most spectra.  

\item Central wavelength is controlled by the grating tilt angle.  A micrometer, with units of thousandths of an inch, is used to set the tilt angle.  Linear equations relate the micrometer value to the approximate central wavelength for each grating, provided in Appendix A.  In the FITS headers, starting wavelength is recorded as CRVAL1 and the pixel scale (\AA\ pix$^{-1}$) is recorded as CD1\_1.

\item FAST has five slits with widths of 1.1, 1.5, 2.0, 3.0, and 5.0 arcsec.  All slits are 3 arcmin long in the spatial direction.  Galaxy and supernova spectra are normally obtained with the 3.0 arcsec slit; stellar spectra are normally obtained with the 1.5 or 2.0 arcsec slits, better matched to typical seeing. Table \ref{tab:fastconfig} summarizes spectral coverage and spectral resolution for the four most common configurations, ordered by the percentage of spectra obtained between 2006 June 2006 and 2019 December.  Most spectra are acquired with the 300 line mm$^{-1}$ grating, the 3.0 arcsec wide slit, and 2$\times$ binning in the spatial direction, a configuration that provides 3475 - 7415 \AA\ spectral coverage at 7.2 \AA\ resolution.  

\item Since 2006 June, FAST's detector has been a backside-illuminated, 2688x512 UA STA520A \#4377 CCD with a blue-enhanced antireflection coating and 15 $\mu$m pixels.  This detector, labeled DETECTOR=`FAST3' in the fits header, has a dark current of less than 1 electron hr$^{-1}$ pix$^{-1}$.  Earlier-generation CCDs had higher dark current and more defects, but identical pixel scale.  All of the thinned detectors exhibit mild 2-3\% amplitude fringing in the red end of the spectrum at $>$6300~\AA.  The spatial scale is always 0.57 arcsec pix$^{-1}$ along the slit.  However, spectra are normally acquired with 2$\times$ or 4$\times$ binning along the spatial dimension.  We tabulate the details of the CCDs that have been used in Table \ref{tab:fastdetect}.  While CCD dimensions have varied, all have had 15$\mu$m square pixels.  The last column indicates the number of processed spectra in the archive obtained with that detector. 
\end{itemize}

\begin{deluxetable}{lcccr}   
\tablecolumns{5}
\tablewidth{0pt}
\tablecaption{FAST Detectors\label{tab:fastdetect}}
\tablehead{
\colhead{Name} & \colhead{Start Date} & \colhead{End Date} & \colhead{Chip Size} & \colhead{Nspec}}
\startdata
CCD & 1994-01-08 & 1994-07-18 & 2688x512  & 3897 \\
 & & & Loral 51222688 & \\
 & & & not thinned & \\
FAST1 & 1994-09-03 & 1998-04-18 & 2720x512 & 28864 \\
 & & & thinned & \\
FAST2 & 1998-04-19 & 2006-06-19 & 2720x512 & 60699 \\
 & & & thinned & \\
FAST3 & 2006-06-20 & 2019-12-22 & 2688x512 & 48041 \\
 & & & thinned &
\enddata
\end{deluxetable}

\subsection{Throughput}

The FAST spectrograph is characterized by high throughput.  What matters for the data is total system throughput:  the product of the telescope mirrors and the spectrograph optics, grating, and CCD.  The total system throughput varies by a factor of 2 or more depending on the state of the optics, but typically peaks around 35\% after primary mirror re-aluminization. 

\begin{figure}
 \includegraphics[width=3.3in]{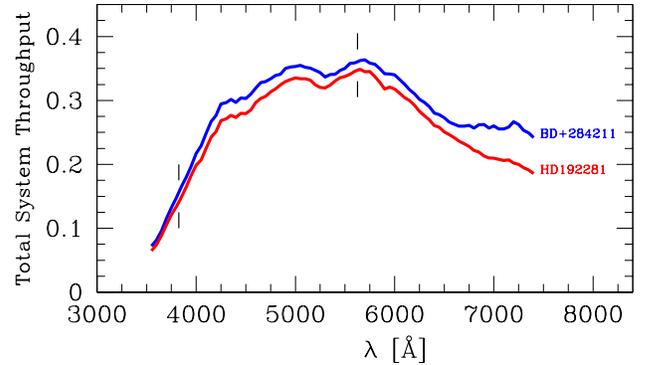}
 \caption{ \label{fig:thruput}
	Total system throughput -- telescope mirrors, spectrograph grating, optics, and CCD -- measured from a hot (BD+284211, blue line) and cool (HD192281, red line) flux standard star on 2014 October 16.  The overall shape reflects the blaze function of the 300 line mm$^{-1}$ grating; second-order light contamination can be seen $>$6500 \AA\ where the blue line diverges from the red line. Black lines mark the wavelengths plotted in Figure \ref{fig:thruputvstime}.}
\end{figure}

Figure \ref{fig:thruput} plots the total system throughput measured from flux standards observed through a 5 arcsec slit (to minimize slit losses) at airmass $<1.04$ on 2014 October 16.  The blue line is measured from a hot star, BD+284211, and the red line is measured from a cool star, HD192281.  The overall shape of the curves is governed by the blaze function of the 300 line mm$^{-1}$ grating.  The low-amplitude oscillations along the curves are real and come from the LLNL Durable Silver coating on the spectrograph collimator and camera mirrors.  The divergence at the red end of the spectrum comes from second-order blue (i.e.\ 3500 \AA) light constructively interfering at first-order red wavelengths (i.e.\ at 7000 \AA).  

\begin{figure}
 \includegraphics[width=3.3in]{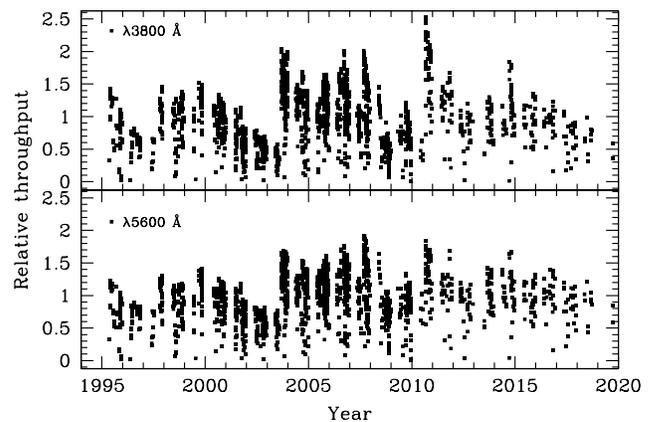}
 \caption{ \label{fig:thruputvstime}
	Relative throughput versus time, measured from the flux of BD+284211 at 3800 \AA\ (upper panel) and 5600 \AA\ (lower panel), normalized to the mean of 25 years of observations.  The upper envelope of points reflects system throughput; low points are meaningless, due to bad pointing or non-photometric conditions.  Jumps in throughput are seen when the primary mirror is re-aluminized or the optics are cleaned. }
\end{figure}

Figure \ref{fig:thruputvstime} plots relative throughput versus time measured from the flux of BD+284211 in two wavelength bands, centered at 3800 \AA\ and 5600 \AA, normalized to the mean of 25 years of observations.  BD+284211 was observed thousands of times in the standard 3 arcsec slit configuration, but not always in photometric conditions.  The upper envelope of points in Figure \ref{fig:thruputvstime} thus provides a measure of relative throughput; the low points are meaningless, due to clouds, bad pointing, and other factors.  Notably, the relative throughput jumps up by a factor of 2 in the red and 4 in the blue when the primary mirror is re-aluminized (every 2 to 4 years) or the corrector optics are cleaned (at more irregular intervals).

\section{Observations} \label{sec:observing}

We provide a brief summary of FAST operations to help users understand the spectra present in the public archive.  We discuss the issue of flux calibration, and then quantify the precision and accuracy of the FAST wavelength calibration.

\subsection{Scheduling}

Observing time is allocated and scheduled on a trimester basis: January-April, May-August, and September-December.  Approved FAST programs are assigned program numbers, stored in the FITS headers as PROGRAM, that identify which spectra belong to which observing program. A program which lasts more than one trimester keeps the same number. Principle investigator name or names are stored in the FITS headers as P.I., with periods. These provide two means of sorting spectra.  

Principle investigators provide the target coordinates for their scientific programs, which are recorded in the FITS header as requested RA and DEC.  Some programs target a single class of sources, such as a galaxy redshift program.  Other programs classify unknown sources, or have target selection that shifts over time, and so accumulate a more diverse set of spectra under a single program number.

FAST observations are acquired in queue mode, not in classically scheduled nights.  Queue scheduling has enabled P.I.s to run long-term monitoring programs as well as target-of-opportunity programs.  Queue scheduling also enables observers at the telescope to optimize target selection and exposure times for the conditions of each night.

\subsection{Observing}

Historically, FAST was mounted on the telescope every month for a $\sim$2 week observing run centered around new moon.  Thus FAST observations are obtained with a nightly and a monthly cadence.  The exception is August, when FLWO closes for maintenance during the worst of the Arizona monsoon weather. 

Observers at the telescope follow a strict protocol.  A complete set of FAST calibrations are obtained every night.  Ten bias frames and five 15 minute dark frames are obtained in each spatial binning.  Twenty-one flat-fields are obtained in each spectrograph configuration.  
All science exposures are immediately followed by a comparison lamp exposure for accurate wavelength calibration.

\subsection{Target Acquisition}

Observers have to visually identify targets on the slit-viewing guide camera, i.e.\ by matching star patterns to finding charts, and manually place targets on the spectrograph slit by eye.  This process can be challenging because the FLWO 1.5m telescope pointing is poor, with an accuracy of about 2 arcmin over most of the sky.  The telescope position recorded in the fits header (TRA, TDEC) thus differs from the requested position (RRA, RDEC).  This discrepancy is normal but uninformative.  The values assigned to RA and DEC in the FITS headers are taken from the requested positions provided by the P.I.'s, and precessed to J2000 if necessary.

FLWO staff observers are very good at identifying targets.  Cross-checks suggest that $>$99.9\% of spectra are of the correct target.  In rare cases, however, the wrong target may be observed.  In other rare cases, the correct object is observed, but the data acquisition system records the wrong coordinate if the telescope was slewed during readout, or if the coordinates were stale from the telescope coming out of a shutdown situation.  Other ``incorrect'' targets are actually real, such as candidate supernovae that turn out to be high proper motion stars superimposed on background galaxies.  Outlier spectra should thus be treated with caution.

\subsection{Flux Calibration}

Every night, observers obtain spectra of red and blue stellar flux standards from \citet{massey88} in every spectrograph configuration used.  The flux standards are labeled with FAST program \#56 Spectrophotometric Standards.  Spectra can be obtained in variable nonphotometric conditions, however.  Thus, flux standards are best used for correcting the spectral continuum shape, and not for determining absolute flux calibration.

Flux calibration was never integrated into the FAST data reduction pipeline, however. The reason is that most FAST spectra have wavelength-dependent flux loss because the observations were made at a fixed position angle of 90$^{\circ}$, not the parallactic angle.  FAST is mounted on a mechanical rotator bearing and can be rotated to any position angle, but the process was time-consuming and rarely used.  The major exceptions were FAST programs \#2 Supernovae and \#56 Spectrophotometric Standards, which set the position angle to the parallactic angle for their targets.  The position angle value is stored in the FITS header as POSANGLE prior to 2002 November 12 and ROTANGLE after 2002 November 12.  

An atmospheric dispersion corrector, installed during the 2010 August shutdown, now makes it possible to more accurately flux-calibrate FAST spectra.  Second-order light still overlaps the red end of the spectrum, however (Figure \ref{fig:thruput}).

Because flux calibration is unnecessary for measuring galaxy redshifts, flux calibration was deferred to the user.  \citet{matheson08} describe how they combine red and blue standards to flux-calibrate FAST spectra for the supernova program.

\begin{figure}
 \includegraphics[width=3.25in]{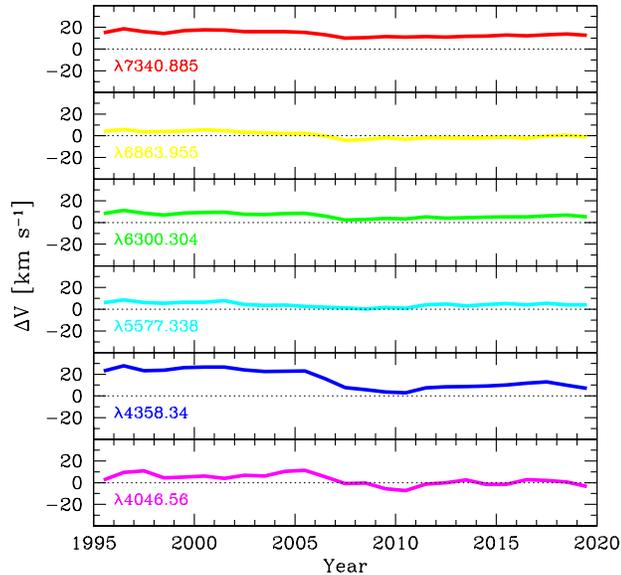}
 \caption{ \label{fig:skylines}
    Weighted-mean velocities of night-sky emission lines relative to their rest-frame velocity (dotted line) versus year of observation.  We conclude that FAST wavelength calibration is accurate at the 10 \kms\ level.}
\end{figure}

\subsection{Wavelength Calibration}

All FAST spectra are wavelength-calibrated using helium-neon-argon comparison lamp exposures obtained immediately after every science exposure.  The only exception to this rule are time-series of short exposures, which can share a single comparison lamp exposure.  

We test the precision of FAST's wavelength calibration with the radial velocities of flux standard stars.  The FAST archive contains over a thousand spectra of stars like Feige34, BD+284211, and HD84937.  We find that a given flux standard exhibits an RMS dispersion of $\pm32$~\kms\ with the 3.0 arcsec slit configuration, and $\pm$12~\kms\ with the 1.5 arcsec slit configurations.  Slit illumination effects likely explain the difference.  Since typical seeing is around 1.5 arcsec, we consider $\pm$12~\kms\ an empirical estimate of radial velocity precision.

Second, we test the accuracy of FAST's wavelength calibration by measuring the radial velocity of night sky emission lines (see Table \ref{tab:skylines}).  The night sky uniformly illuminates the slit and is at rest with respect to the telescope.  We measure velocity with the cross-correlation program EMSAO \citep{kurtz98} for 27,583 spectra obtained with the 300 line mm$^{-1}$ grating and $>$900~s exposure times.  We clip $>$4$\sigma$ outliers, because cosmic rays were not removed from the sky spectra, and compute weighted mean velocities by year for each line.  We plot the results in Figure \ref{fig:skylines}.

Figure \ref{fig:skylines} reveals both wavelength- and time-dependent trends.  Prior to the FAST3 CCD upgrade in 2006 June, the night sky lines are systematically shifted in velocity from +5 \kms\ at $\lambda5577$ \AA\ to about +20 \kms\ at the red and blue ends of the spectrum.  We attribute the systematic offset to the difference in how the helium-neon-argon calibration lamp illuminates the spectrograph compared to the night sky.  The accuracy improves after 2006 June, to a weighted mean over all lines of $+5.6\pm0.1$ \kms.  We conclude that the FAST wavelength calibration has an effective precision and accuracy at the 10 \kms\ level.

\begin{deluxetable}{cccc}   
\tablecolumns{4}
\tablewidth{0pt}
\tablecaption{Weighted-Mean Velocity of Night-Sky Lines\label{tab:skylines}}
\tablehead{
     \colhead{Line} & \colhead{$\lambda$} & \colhead{RV $_{{\rm pre-}2006.5}$} & \colhead{RV $_{{\rm post-}2006.5}$} \\
                    & (\AA)               & (\kms)                    & (\kms)  }
\startdata
Hg{\sc i}           & 4046.56  & $ +7.1$ & $ -1.1$ \\
Hg{\sc i}           & 4358.34  & $+24.3$ & $ +7.5$ \\
Hg{\sc i}           & 5460.74  & $+22.8$ & $+13.8$ \\
$[$O{\sc i}$]$     & 5577.338 & $ +5.5$ & $ +2.9$ \\
$[$O{\sc i}$]$     & 6300.304 & $ +8.3$ & $ +4.1$ \\
$[$O{\sc i}$]$     & 6363.776 & $ +7.4$ & $ +4.6$ \\
OH7-2Q1(1.5) & 6863.955 & $ +3.7$ & $ -2.4$ \\
OH8-3P1(3.5) & 7340.885 & $+16.2$ & $+11.4$ \\
\enddata
\end{deluxetable}

\section{Data Reduction Pipeline} \label{sec:pipeline}

FAST spectra have been processed with essentially the identical IRAF-based data reduction pipeline \citep{tokarz1997} since first light in 1994 January.  In the first decade of operation, however, only data obtained with the ``standard'' DISPERSE=300, TILTPOS=590, APERTURE=3 configuration were processed.  P.I.'s were expected to process the other $\sim$10,000 spectra obtained in nonstandard configurations.

Beginning in 2006 June (with the FAST3 CCD upgrade), we updated the pipeline to work with all spectrograph configurations, and have processed every FAST observation.  As of 2019 December 31, the SAO OIR Telescope Data Center has processed 156,899 of the 166,979 total science spectra acquired by FAST.

\begin{figure}
\includegraphics[width=3.25in]{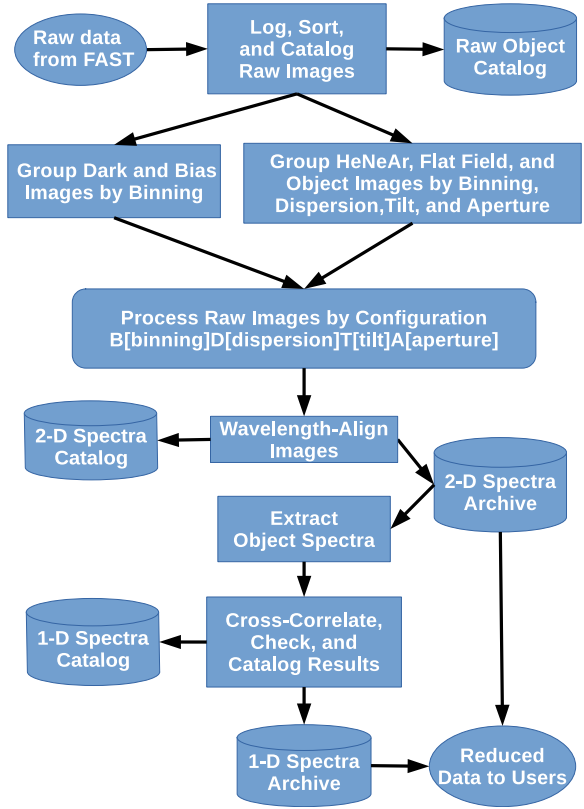}
\caption{ \label{fig:workflow}
	Workflow of the daily processing, archiving, and distribution of FAST spectra.}
\end{figure}

We process, archive, and distribute FAST data on a daily basis.  Figure \ref{fig:workflow} illustrates the workflow.  Each morning, a cron job automatically transfers raw data from Mt.\ Hopkins to a local archive at SAO.  We use WCSTools \citep{mink2002} to read the FITS headers, and Unix shell scripts to sort the data by spectrograph configuration.  Data from each configuration is then run through the IRAF data reduction pipeline.  The results are cataloged, archived, and distributed to the appropriate P.I.s.  The final data products are linearized, wavelength-calibrated two-dimensional images and one-dimensional extracted spectra.  

We use IRAF tasks to perform all data reduction steps, and run them using a locally written IRAF script named ROADRUNNER \citep{tokarz1997}.  The IRAF CCDPROC task performs bias correction and flat fielding.  Dark subtraction is done only for data obtained prior to June 2006 with earlier-generation detectors.  ROADRUNNER looks up the appropriate line list and master comparison file for a given spectrograph configuration.  IRAF's REIDENTIFY task refits the wavelength solution to each comparison lamp image using a low-order polynomial.  We perform 16 unique solutions along the spatial direction, in 11.4 arcsec (20 unbinned pixel) intervals, to deal with any rotation or distortion of the slit image on the detector.  After 2006 June, each two-dimensional science exposure is then transformed by the polynomial fits to be linear in wavelength, and written to a FITS image file with a 'TF' appended to the object name. 

Note that, unlike the Sloan Digital Sky Survey (SDSS) pipeline, we do not apply the barycentric velocity correction to the wavelength solution.  Rather, the barycentric velocity correction is stored in the FITS header as BCV and applied when measuring an astronomical target's radial velocity (see below).  This choice means that telluric night-sky lines correctly appear at their rest-frame wavelengths in the spectra.

We use a second locally written IRAF script named BEEPBEEP to extract the spectra.  It begins by running our IRAF task FINDALL to locate the brightest object on the slit and select a background region for extraction, working around any other sources it finds on the slit.  The selection can be run interactively and to make sure the correct object is selected.  The background and object regions are saved in the FITS header.  Next, the object is optimally extracted into a one-dimensional spectrum using IRAF's APALL task.  If a second object appears on the slit, it can be extracted in a second pass.
From 1994 to 2006, object and calibration spectra were extracted from unaligned processed images, dispersion functions were fit to the closest HeNeAr spectra and object spectra were assigned dispersion functions based on those solutions.

Each extracted spectrum is written into a multiband FITS file that provides the extracted object, sky, object+sky, and, after 2006 June, the variance spectrum.  The spectra are binned into 2681 pixels with a linearized wavelength solution given in the FITS header (the CD matrix parameters). The tables in Appendix C describe describe the full list of FITS header keywords inserted by the data reduction pipeline.  Since June 2006, the multi-band FITS files have dimensions of [2681,1,4].

\begin{figure}
\includegraphics[width=3.25in]{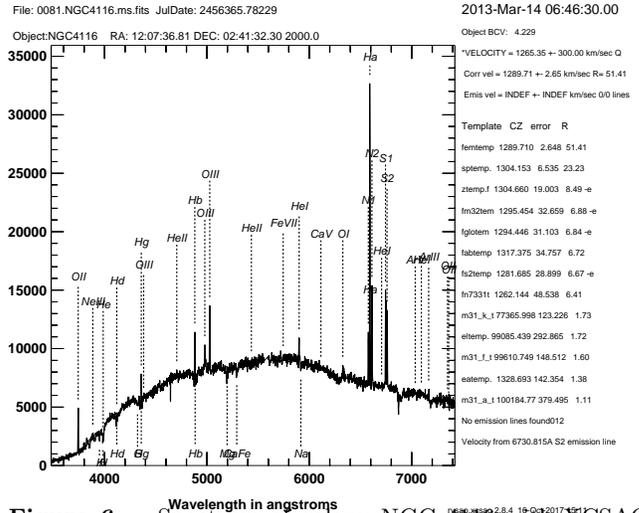}
\caption{ \label{fig:xcsao}
    Spectrum of galaxy NGC 4116 with XCSAO cross-correlation results.}
\end{figure}

\begin{deluxetable*}{rrcll}
\tablecolumns{5} \tablewidth{0pt}
\tablecaption{Largest Programs in the Public Archive (2019 December 31)\label{tab:programs}}
\tablehead{ \colhead{N$_{\rm spectra}$} & \colhead{N$_{\rm unique}^*$} & \colhead{Prog.~\#} & \colhead{Program Title} & \colhead{P.I.} }
	\startdata
17019&  103 & 56 & Spectrophotometric Standards & All \\
11913&  457 & 57 & Velocity Standard & All \\
8132 & 1720 &  2 & Supernovae & Kirshner  \\
8071 &   95 &  6 & AGN Monitoring & Wilkes \\
7622 & 6803 & 68 & 2MASS Redshift Survey & Huchra \\
6938 & 6400 & 64 & Cluster Infall Regions & Geller \\
5764 & 5200 & 141 & 2MASS Redshift Survey & Falco \\
5301 & 4482 &  3 & 15R Redshift Survey & Geller \\
5013 & 1911 & 178 & Low Mass White Dwarfs & Brown \\
4911 & 3302 &  1 & CfA Redshift Survey & Huchra/Geller \\
4029 & 3472 & 112 & Pre-Main Sequence Stars in Orion & Calvet \\
2936 & 2795 & 59 & X-ray Groups & Mahdahvi \\
2927 & 2449 & 157 & IPHAS & Steeghs \\
2263 &    2 & 12 & Symbiotic Stars & Kenyon \\
2128 & 1045 & 89 & Ae/Be Stars & Calvet \\
2050 & 1413 & 60 & CfA Redshift Survey Completion & Geller \\
	\enddata
\tablenotetext{*}{Defined as targets with unique OBJECT names separated by $>$10 arcsec, excluding twilight-sky spectra.}
\end{deluxetable*}

\subsection{Cross-Correlation and Quality Control}

The goal of many FAST observing programs is to measure the radial velocities of galaxies or stars.  Thus, we cross-correlate each spectrum as a final quality control step.

We perform cross-correlation with the package RVSAO described in detail by \citet{kurtz98}.  Briefly, we subtract the continuum, apodize, and Fourier filter each spectrum.  We then cross-correlate the full spectrum against a library of stellar and galaxy templates, rebinned onto a common log-wavelength scale.  We estimate the velocity and velocity error from the peak and width of the cross-correlation peak following \citet{tonry1979}.

We present the cross-correlation templates in Appendix B, and caution that some spectral types of objects -- quasars, supernovae, M dwarfs, helium stars, continuum objects -- have no valid template and/or can latch on to a bogus template.  We use QPLOT to visually inspect and validate the velocity of each spectrum.  

Figure \ref{fig:xcsao} shows the example of NGC~4116 with its best-fit emission and absorption lines labeled. The QPLOT task allows us to remove cosmic rays, which can confuse emission line fits, and recorrelate if necessary.  We then set the quality flag to approve (VELQUAL=``Q'') or disapprove (VELQUAL=``X'') the cross-correlation.

The full list of cross-correlation keywords is presented in Appendix C; the most important keywords for the user are VELOCITY, the best-fit radial velocity in km~s$^{-1}$, CZXCERR, the velocity error in km~s$^{-1}$, BESTTEMP, the best-matching template, and CZXCR, the $r$ statistic (analogous to signal to noise) of the fit.  These values are a guide to the nature of an object (BESTTEMP, VELOCITY) and the quality of its spectrum (CZXCR, CZXCERR).

Scientists typically perform follow-up analysis optimized for their sources.  Published velocities thus supersede the velocities recorded in our FITS headers.  

\subsection{Data Archiving and Distribution}

Processed 1D and 2D spectra are normally distributed to P.I.s within one working day of acquisition, using a shell script that sends email notifications with download instructions.  Both raw and processed files are archived for permanent storage.  Files are stored first to a local RAID-array archive in Cambridge, MA, and are mirrored quarterly to an off-site Smithsonian computing facility in Herndon, VA.  Copies are also retained at FLWO for several years, to ensure there is always a copy at multiple distinct sites.

Metadata used by the public archive are ingested and stored in a PostgreSQL database residing in Cambridge and on our cloud server that hosts the \url{http://oirsa.cfa.harvard.edu} public interface.  The public archive is described next.

\section{The Public Archive}

The CfA Optical Infrared Science Archive (OIRSA) was released by the SAO Telescope Data Center in early 2016.  The goal was to make a complete set of \emph{processed} spectra, obtained from optical/infrared spectrographs on the CfA's ground-based telescopes, available to the public.  

OIRSA uses the German Astrophysical Virtual Observatory Data Center Helper Suite (DaCHS) \citep{2014A&C.....7...27D}, a publishing infrastructure for the Virtual Observatory. DaCHS exploits the open-source PostgreSQL object-relational database management system\footnote{\url{https://www.postgresql.org/}} with specific extensions to provide flexible data search and access mechanisms to astronomical data \citep{2004ASPC..314..225C}. The DaCHS access mechanisms are built on top of an existing PostgreSQL database schema and operate on an SQL view created from joining several data tables. Our database schema implements our own proprietary data model, which we reuse with minor modifications across the public and private data archives for different instruments and also for operations of the Binospec \citep{2019PASP..131g5004F} and MMIRS \citep{2012PASP..124.1318M} spectrographs at the 6.5-m MMT telescope.

DaCHS data access services for the OIRSA archive implement several International Virtual Observatory Alliance (IVOA) standards: (i) the Simple Spectral Access Protocol \citep{2012ivoa.spec.0210T} providing access to one-dimensional extracted spectra; (ii) Observation Data Model Core Components, its Implementation in the Table Access Protocol \citep{2011ivoa.spec.1028T} providing a more generic and flexible query interface to the metadata search including the cross-match via table upload; (iii) Registry Interfaces \citep{2018ivoa.spec.0723D} that tell the Virtual Observatory about the existence of OIRSA data access services and make it discoverable in client applications such as {\sc topcat} \citep{2005ASPC..347...29T}.

The OIRSA archive debuted with over 290,000 processed 6.5 m MMT Hectospec spectra, homogeneously reduced with the latest data reduction pipeline and linked to published papers.  The Hectospec archive is updated yearly and includes every spectrum older than 10 years old.  Second to be added were over 250,000 single-order echelle spectra, each with wavelength coverage of 45 \AA\ and spectral resolution of about 36,000 centered at 5187 \AA.  The echelle spectra were obtained with cross-dispersed echelle spectrographs using 2 x 936-pixel Reticon detectors on the FLWO 1.5-meter Tillinghast telescope, the 4.5 m-equivalent Multiple Mirror Telescope, and the 1.5-meter Wyeth telescope at Oak Ridge Observatory in Harvard, Massachusetts \citep{mink2016}.  The echelle spectra were primarily obtained for radial velocity studies of bright stars, and the uniformly processed spectra provide a valuable resource in combination with Gaia proper motion and parallax measurements.

DaCHS provides a simple web-based query form, which allows one to search and download spectra in bulk and which actually queries TAP and SSAP services in the backend.  For example, a list of up to 280 target positions along with search radii can be queried with the web form; OIRSA returns the resulting batch of spectra in a tarball file.  Much larger queries can be made with Virtual Observatory tools like TopCat \citep{2005ASPC..347...29T}.  
Thanks to the implementation of IVOA standards, the OIRSA FAST Archive can be queried in a batch mode using scripts in several programming languages (Python, Shell, IDL, Java) using Application Programming Interface libraries.  

\section{Recommendations for Use}

OIRSA currently contains 141,531 processed FAST spectra obtained between 1994 January and 2019 December.  We have removed many of the completely failed observations, such as saturated spectra or spectra with zero counts.  However, we have erred on the side of retaining everything else.  The archive contains over 5,500 twilight sky spectra, for example, observations that can sometimes point at real objects although only twilight sky is visible.

A good rule of thumb is that spectra with CZXCR$>$3 are mostly reliable; spectra with CZXCR$<$3 are mostly noise.  Fully 87.5\% of the reduced FAST spectra have VELQUAL=Q and CZXCR$>$3, spectra that should be ready to use.

A generous search radius should be used if cross-matching the FAST archive by position.  We record RA and DEC in epoch J2000, but the equinox on which the coordinates are based is unknown.  Galaxy coordinates that come from photographic plates have $\sim$10 arcsec coordinate uncertainties \citep{falco99}.  Stellar coordinates that come from modern surveys like SDSS have $<$1 arcsec uncertainties but, unlike galaxies, can have significant proper motion.  

A powerful approach for using FAST spectra is to start with published, well-defined surveys.  Table \ref{tab:programs} lists the largest programs in the public archive:  all programs with $>$2000 spectra as of December 31, 2019.  The median visual magnitude of targets in the FAST archive is around 15 mag; the faintest point sources are 18 mag.

\subsection{Signature FAST Programs}

Signature programs include the CfA Redshift Survey \citep{geller89}, Updated Zwicky Catalog \citep{falco99}, and the 2MASS Redshift Survey \citep{huchra12}.  A number of the redshifts for those surveys were measured with FAST.

Redshift surveys made entirely with FAST observations include the 15R Redshift Survey \citep{carter01}, the Cluster and Infall Region Nearby Survey \citep{rines03}, and the X-ray emitting nearby galaxy groups program \citep{mahdavi04}.

One of the largest programs is the CfA Supernova Program, which obtained over 8000 spectra for bright $V<18$ mag supernovae over the past quarter century.  The public archive contains the spectra but not the final reductions published by the CfA Supernova Program, so we refer the user to their papers \citep{matheson08, blondin12, hicken17}. 

Other major holdings include Galactic plane surveys of young emission line stars \citep{hernandez05, raddi15}, long-term monitoring programs of active galactic nuclei \citep{trichas12} and of symbiotic star binaries \citep{kenyon2016, cuneo18}, and time-series spectroscopy of low-mass white dwarf candidates \citep{brown10}.  A few programs are missing from the archive because their spectra were obtained prior to 2006 in nonstandard configurations.  The largest missing program is the Century Survey Galactic Halo Project (PROGRAM=122) that measured radial velocities for over 4,000 blue horizontal branch star candidates \citep{brown08}.

Selecting spectra by PROGRAM or P.I. does not guarantee a complete set, however.  Thus, we provide tar files of spectra for some major programs on the public archive web page.

\section{CONCLUSIONS}

Our hope is that making reduced FAST spectra easily accessible will enable scientists to make good use of the spectra for decades to come.  A surprising number of 16th magnitude galaxies in SDSS lack redshifts due to the fiber-positioning constraints of the SDSS spectrograph \citep{lazo18}, for example.  Some of those galaxies are found in the FAST archive.  Gaia likely provides parallax and proper motions for every star that FAST has ever observed, but most of the hot stars lack radial velocities, velocities that can be found in the FAST archive.  We encourage users to explore the spectra collected with FAST over the past quarter century.  The OIRSA will add processed spectra taken with other CfA instruments in future years.

\acknowledgments{The data in this paper were taken using the FAST spectrograph by a number of observers, the most prolific of whom were Perry Berlind (70,891 spectra), Mike Calkins (38,543 spectra), Jim Peters (4345 spectra), Ken Rines (3953 spectra), John Huchra (3762 spectra), Warren Brown (3759 spectra), and Lucas Macri (2957 spectra). I.C. is supported by the Telescope Data Center at SAO and also acknowledges the Russian Science Foundation grant 19-12-00281.}

\facility{FLWO:1.5m}

\appendix

\section{Grating Tilt Equations}

The following relations link grating tilt, set by the micrometer TILTPOS value, to the approximate central wavelength (CW) in units of \AA.  The 1200 line mm$^{-1}$ grating has two equations because the micrometer must be used with a spacer to tilt the grating to blue wavelengths.

\begin{eqnarray}
  300~{\rm l~mm^{-1}} ~~ & {\rm CW} = 13.2*{\rm TILTPOS} - 2400~{\rm \AA} \\
  600~{\rm l~mm^{-1}} ~~ & {\rm CW} = 6.61*{\rm TILTPOS} + 1600~{\rm \AA} \\
  1200~{\rm l~mm^{-1}} ~~ & {\rm CW} = 3.29*{\rm TILTPOS} + 5200~{\rm \AA} \\
  {\rm spacer} + 1200~{\rm l~mm^{-1}} ~~ & {\rm CW} = 3.51*{\rm TILTPOS} + 3850~{\rm \AA}
\end{eqnarray}

\section{FAST Cross-correlation Templates}

Our quality control check uses one of two sets of cross-correlation templates listed in Table \ref{tab:fasttemps}, a total of 37 spectra assembled at different times for different programs. Most of the templates were made empirically, by combining high signal-to-noise observations of objects of known velocity using the IRAF RVSAO SUMSPEC task \citep{mink2013} or using our IRAF singular value decomposition task SVDFIT \citep{kurtz2000} when a large number of spectra were combined.  Emission line templates were generated by the LINESPEC task in the RVSAO package \citep{kurtz98}. All of the cross-correlation templates are distributed with the RVSAO package \citep{kurtz1999}.  Characteristics of all of these templates are given in Table \ref{tab:fasttemps}. 

\begin{figure*}
\gridline{\fig{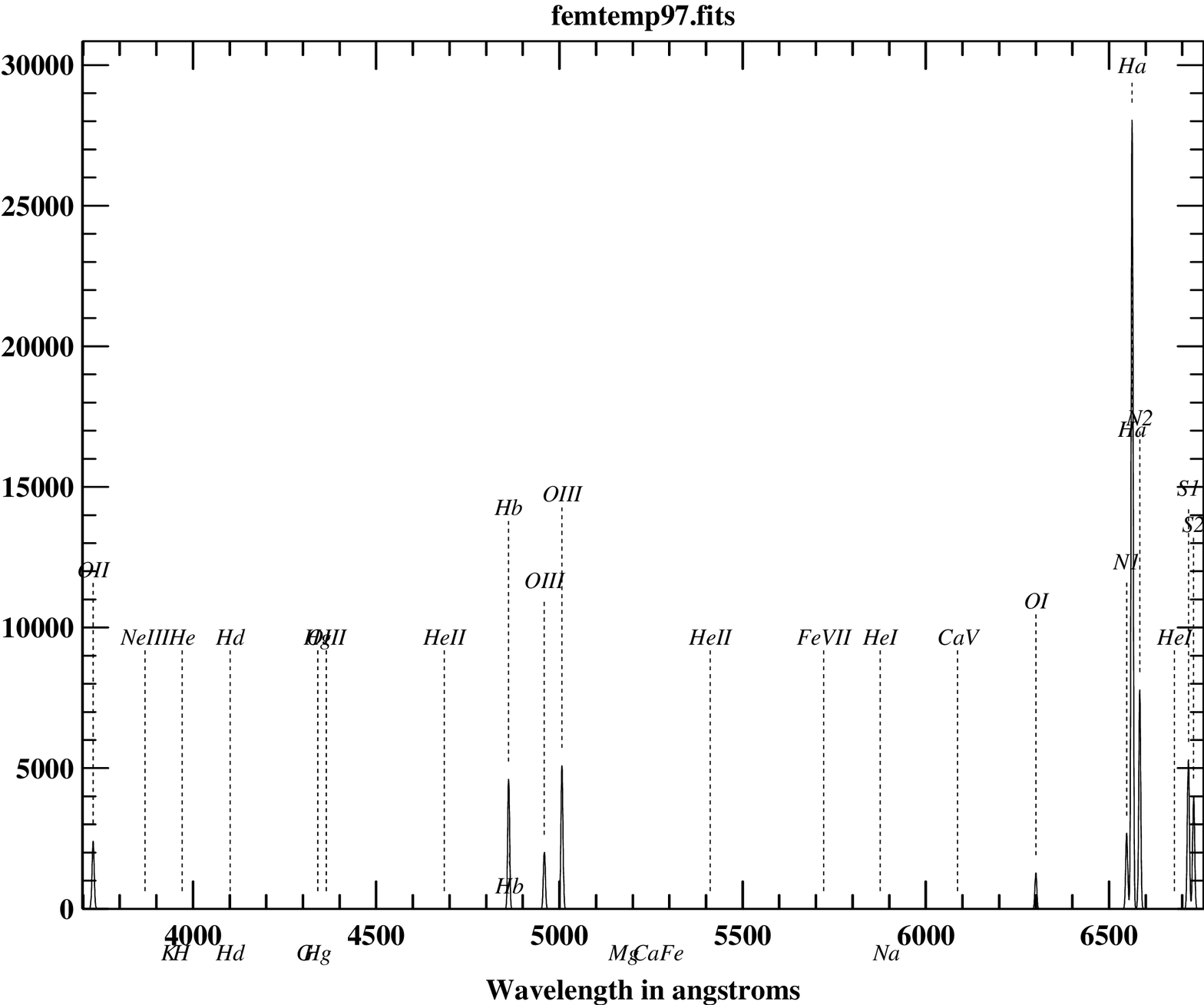}{0.26\textwidth}{a. femtemp97 -- galaxy emission line template (39054).}
          \fig{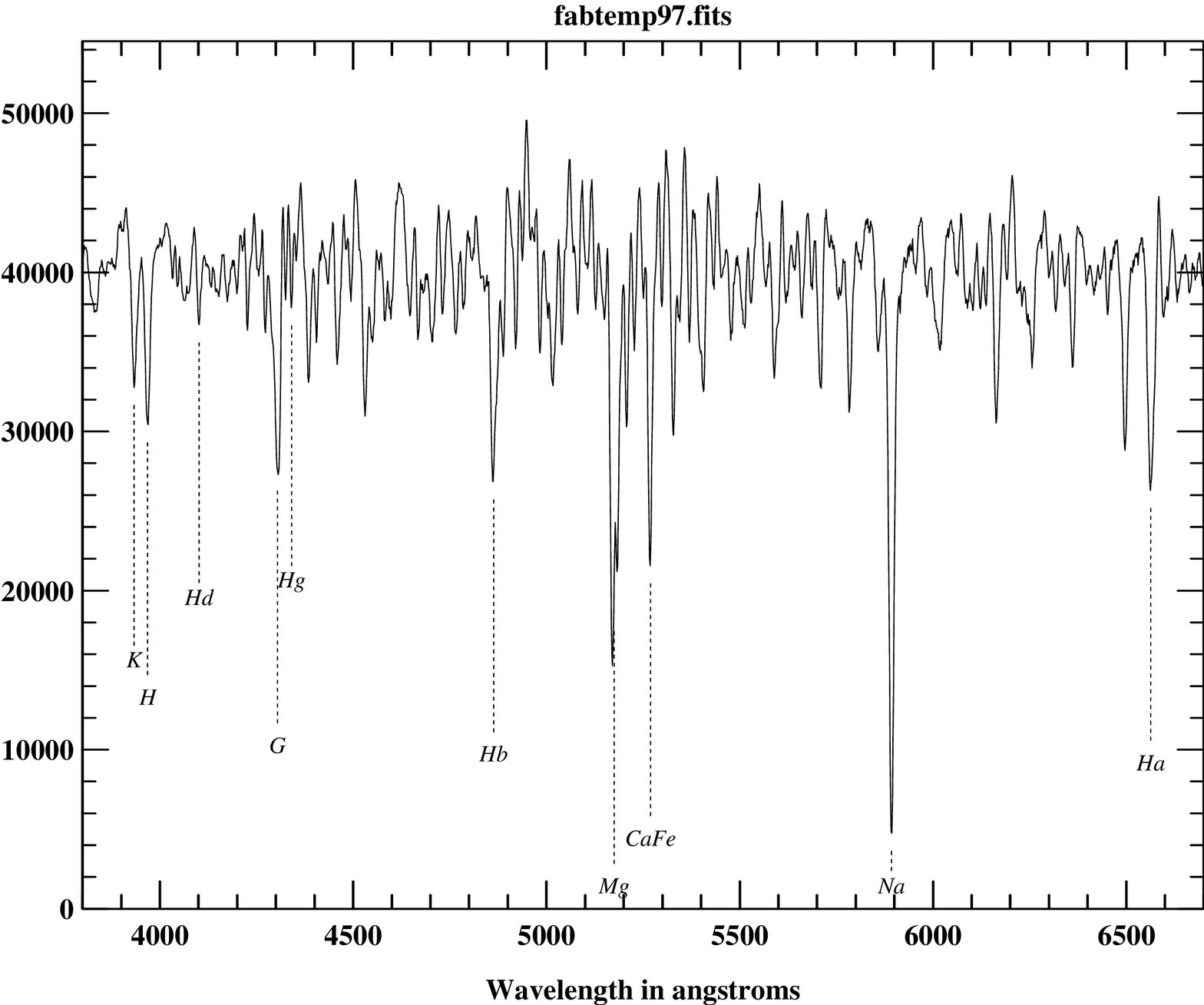}{0.26\textwidth}{b. fabtemp97 -- galaxy absorption line template (26639).}
          \fig{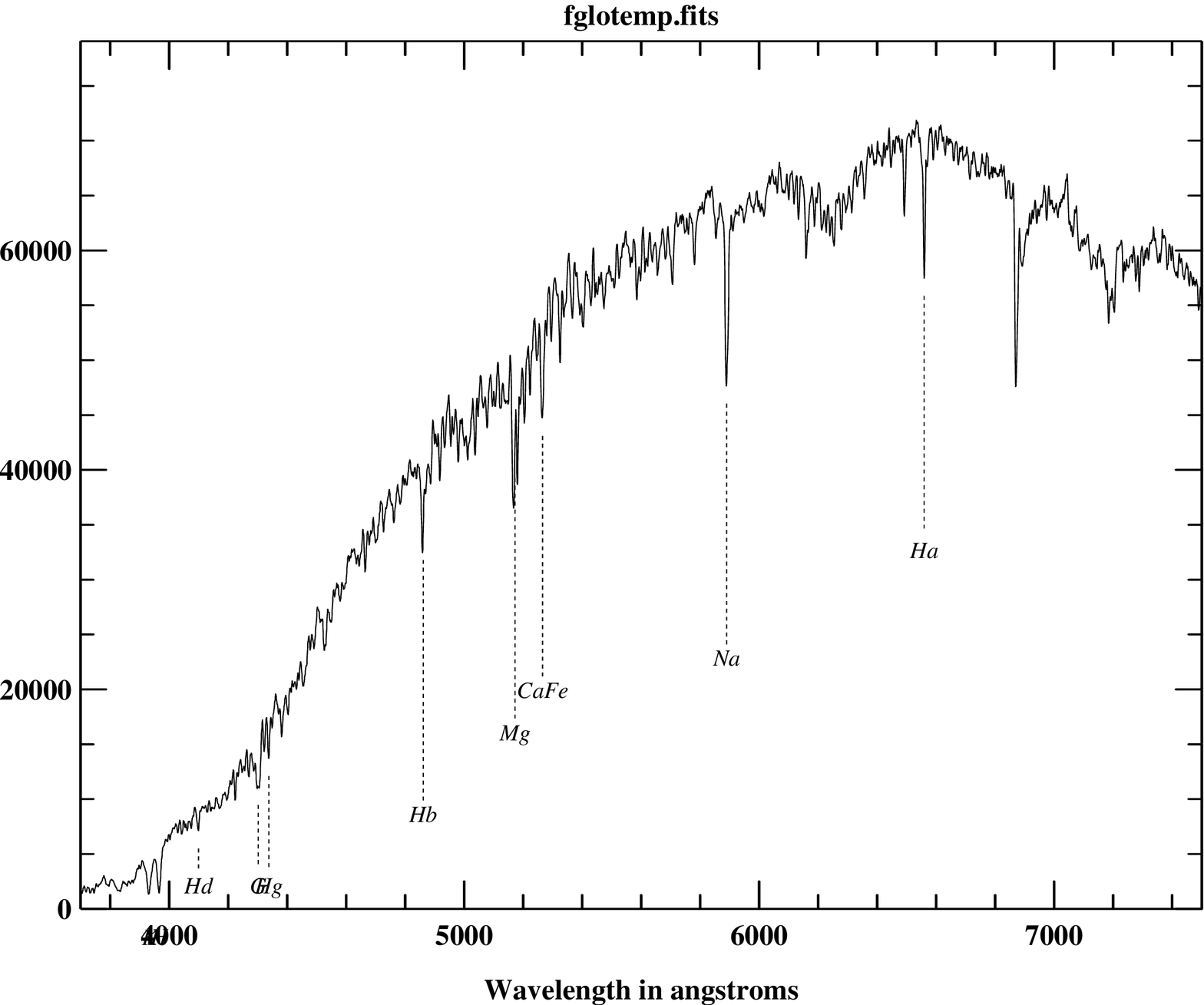}{0.26\textwidth}{c. fglotemp -- globular cluster template (19920).}
          }
\gridline{\fig{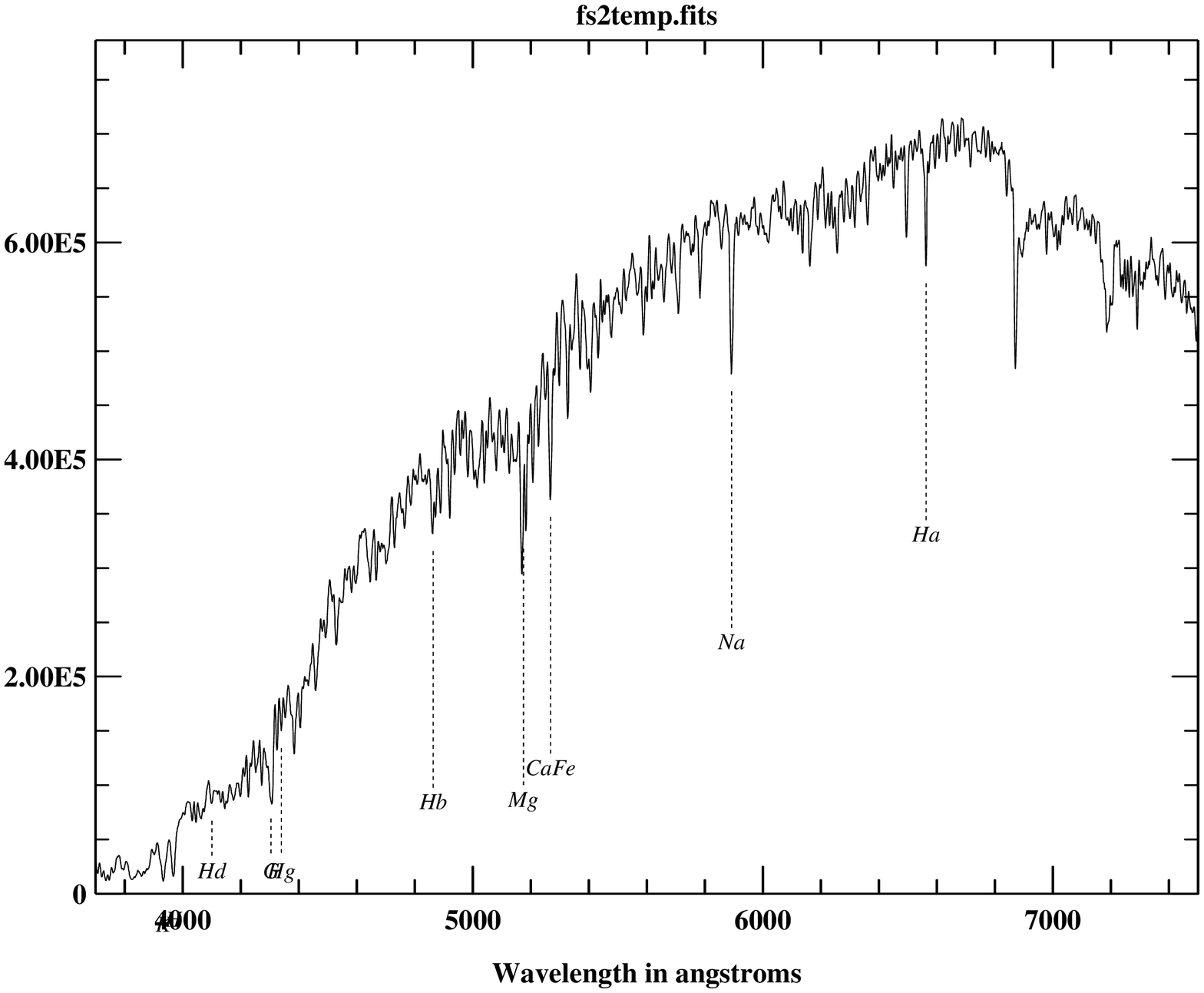}{0.26\textwidth}{d. fs2temp -- stellar K-star template (14486).}
          \fig{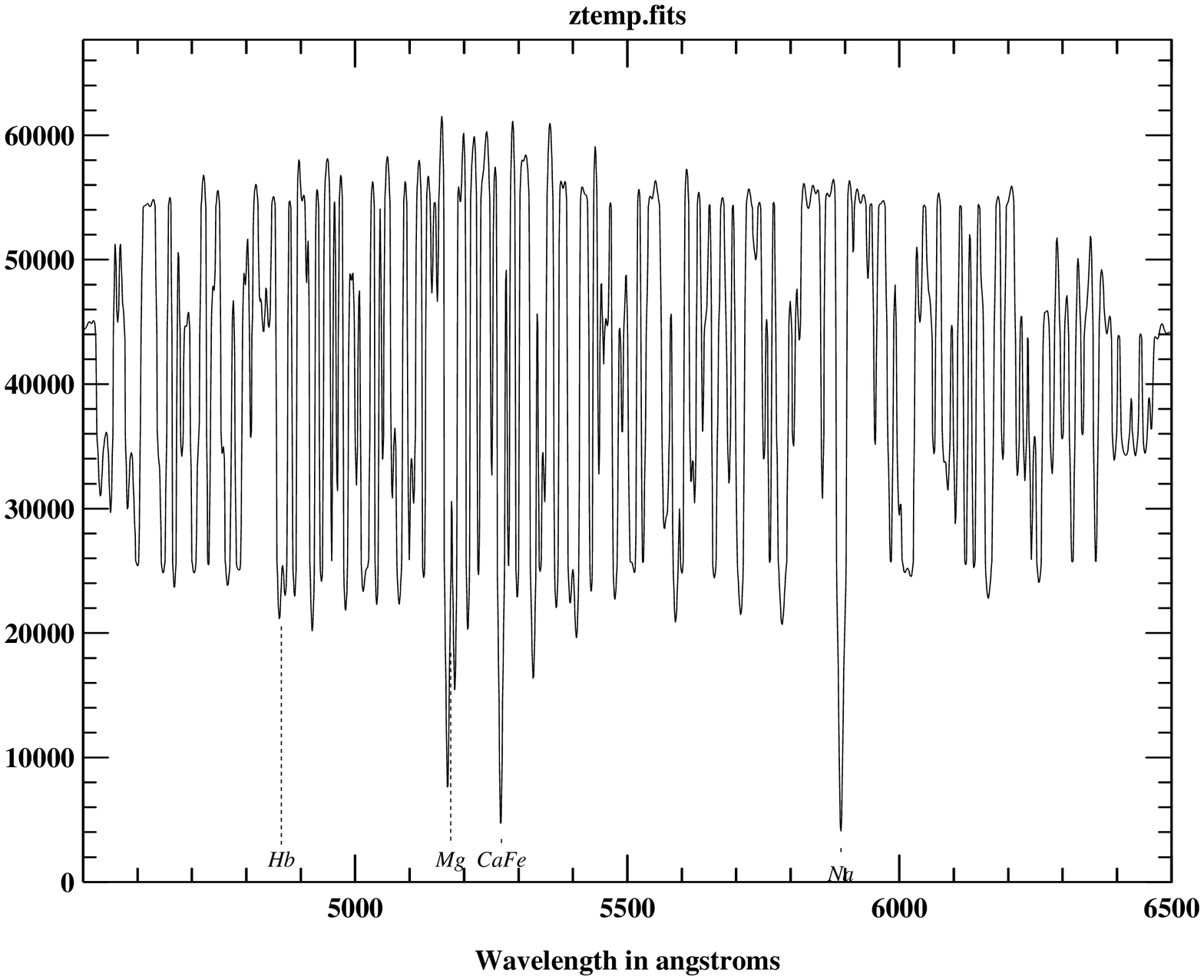}{0.26\textwidth}{e. ztemp -- Z-Machine galaxy absorption line template (10629).}
          \fig{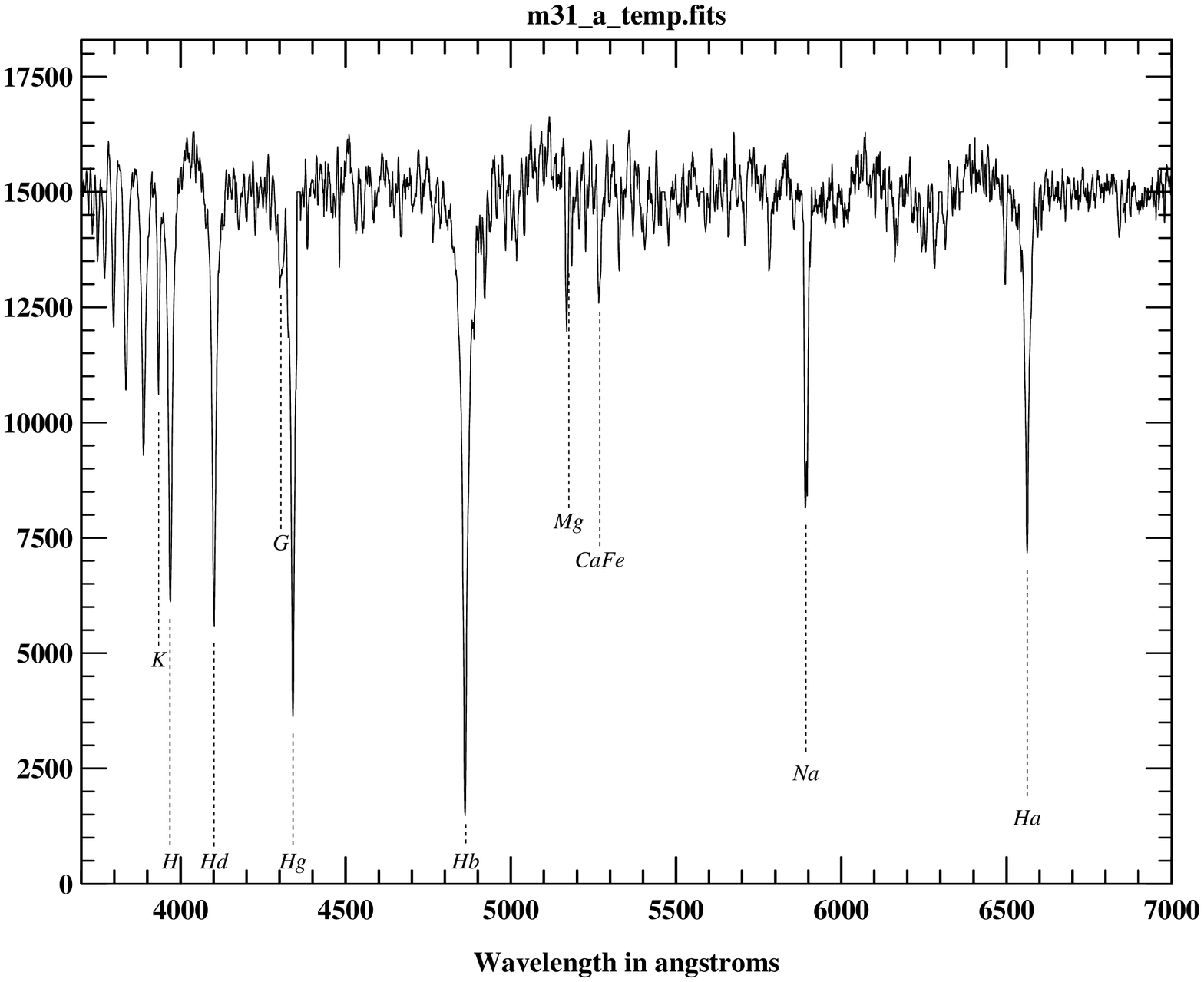}{0.26\textwidth}{f. m31\_a\_temp -- halo A-star template (7022).}
          }
\caption{\label{fig:commontempl}Six most-used cross-correlation templates in the FAST reduced spectra archive (number as of Dec 31, 2019).}
\end{figure*}

\begin{figure*}
\gridline{\fig{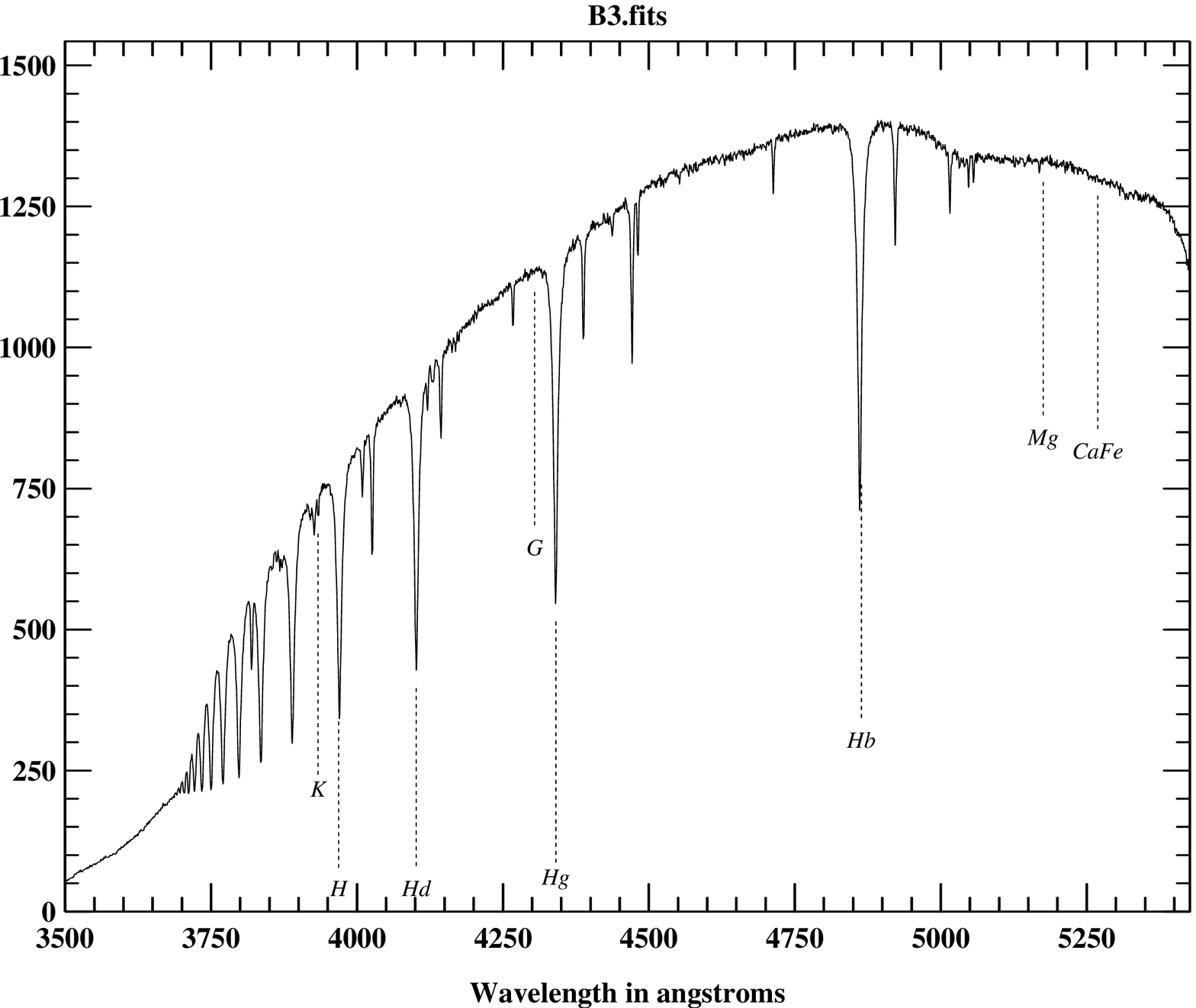}{0.26\textwidth}{a. B3 template.}
          \fig{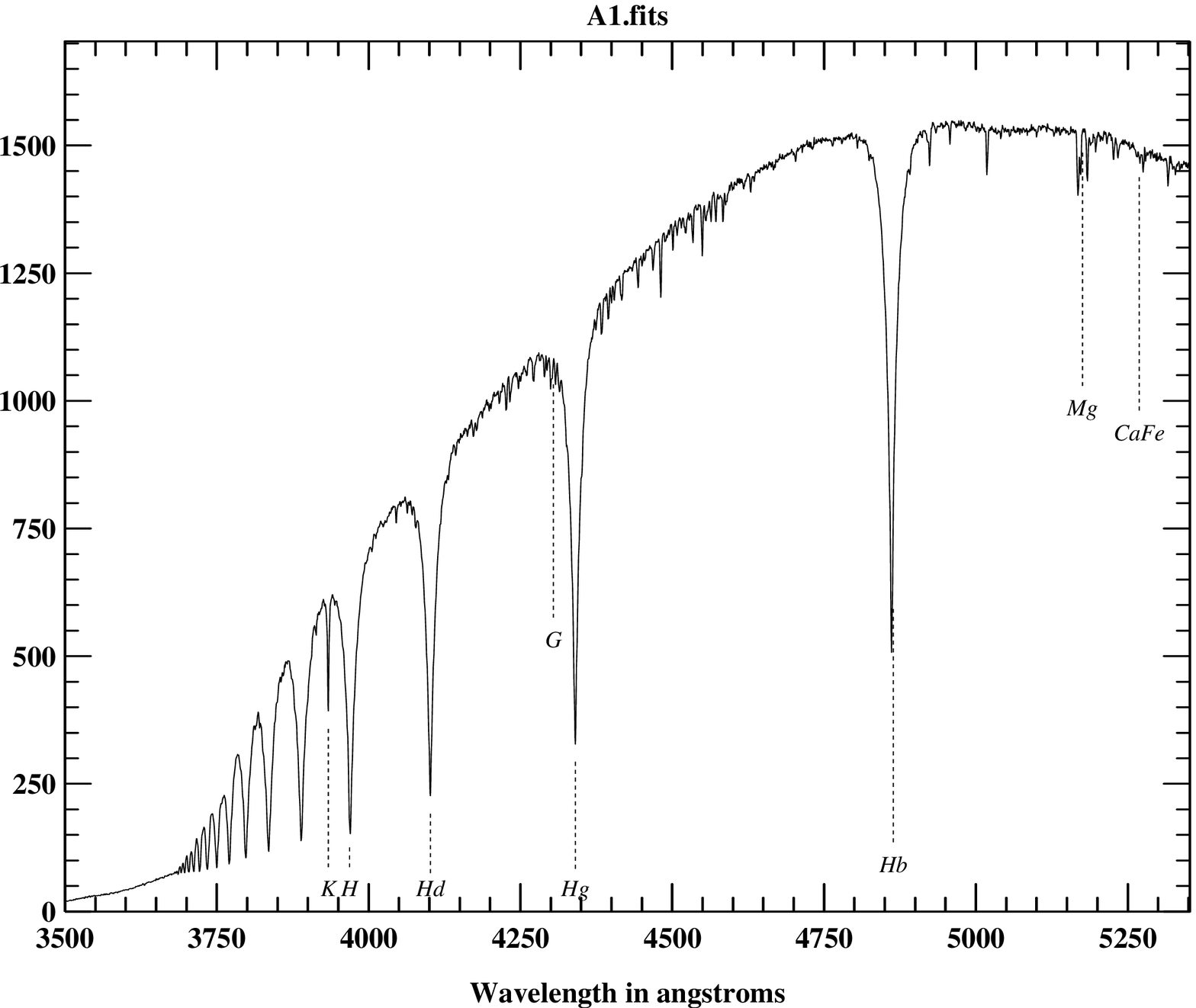}{0.26\textwidth}{b. A1 template.}
          \fig{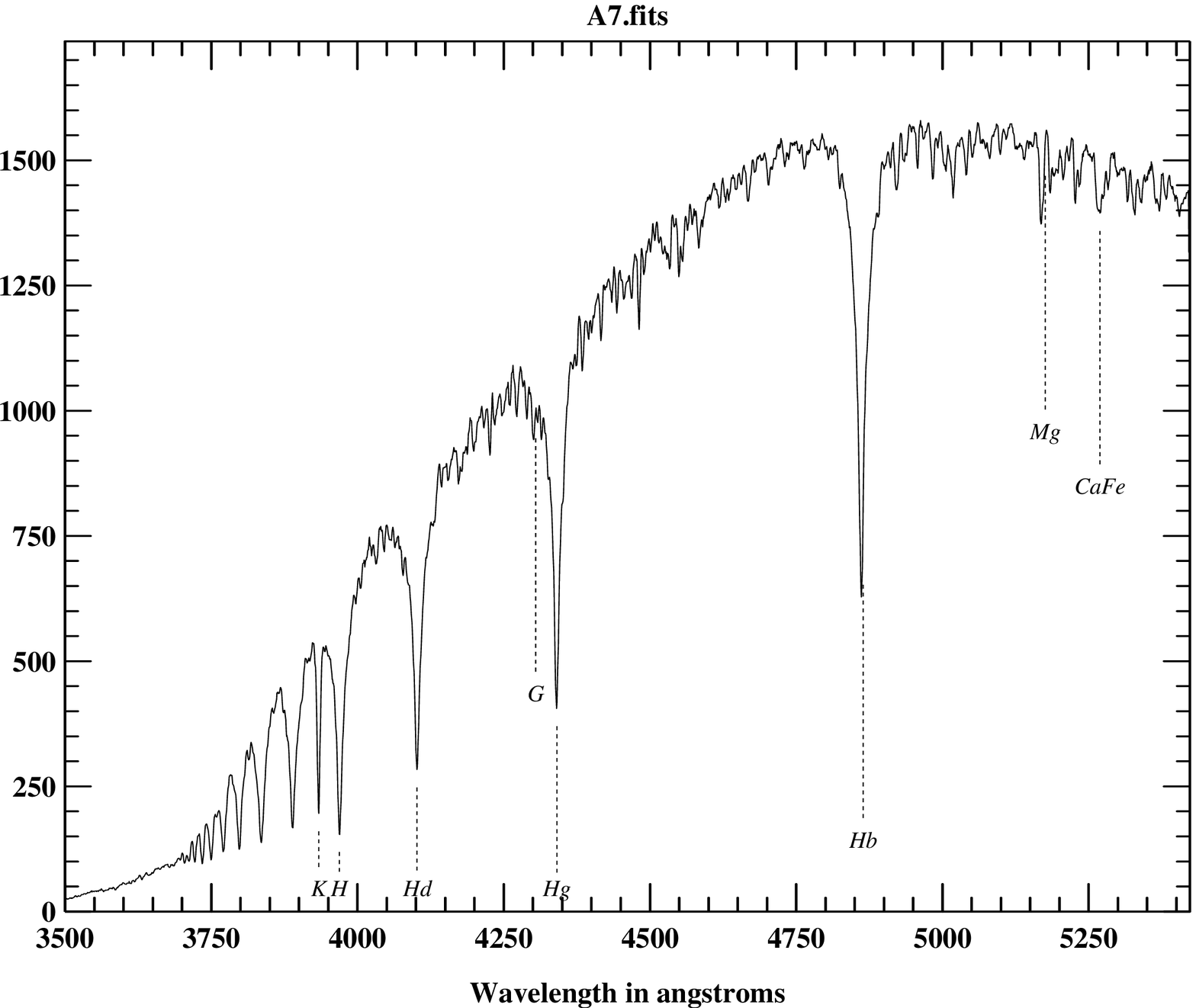}{0.26\textwidth}{c. A7 template.}
          }
\gridline{\fig{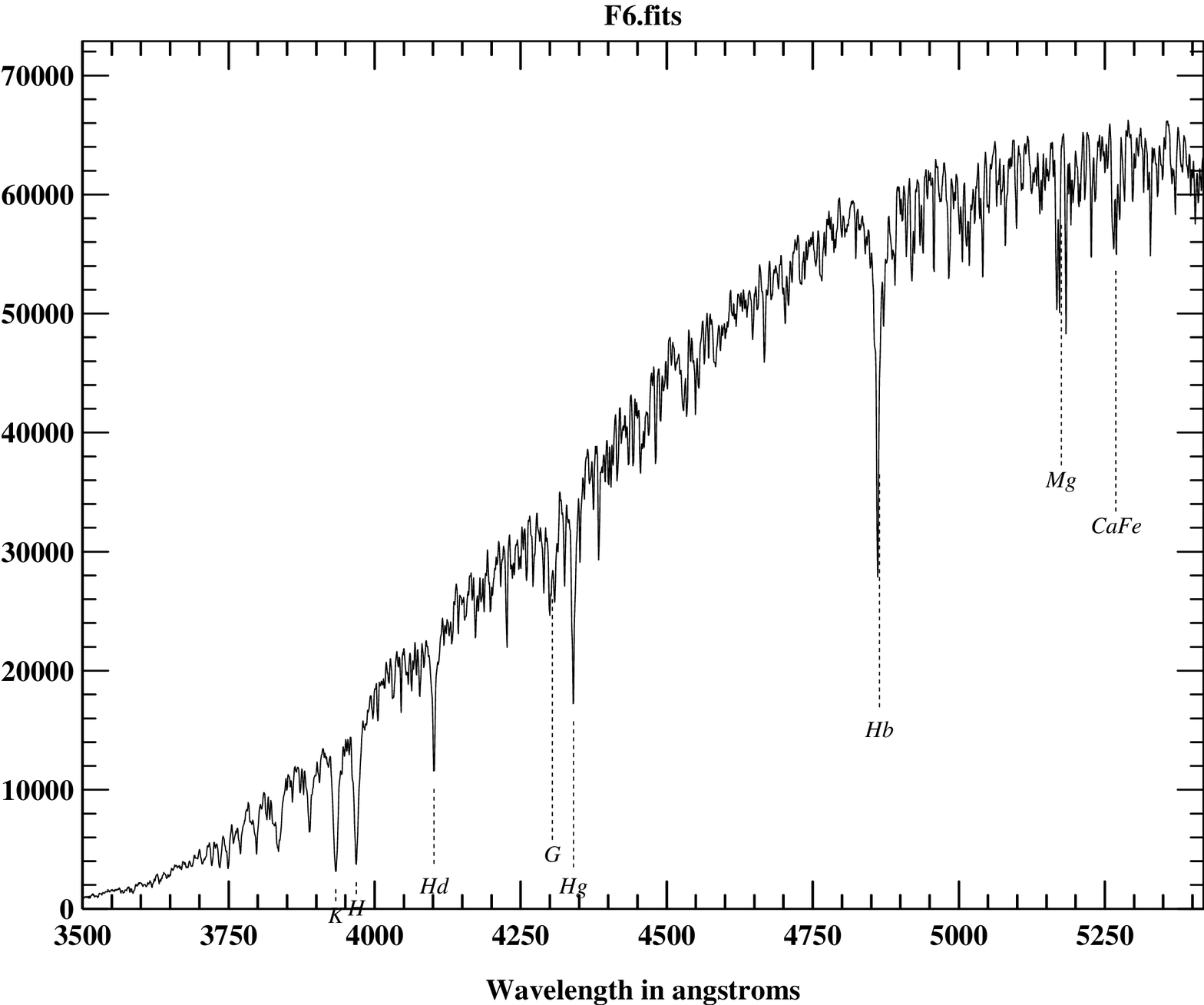}{0.26\textwidth}{d. F6 template.}
          \fig{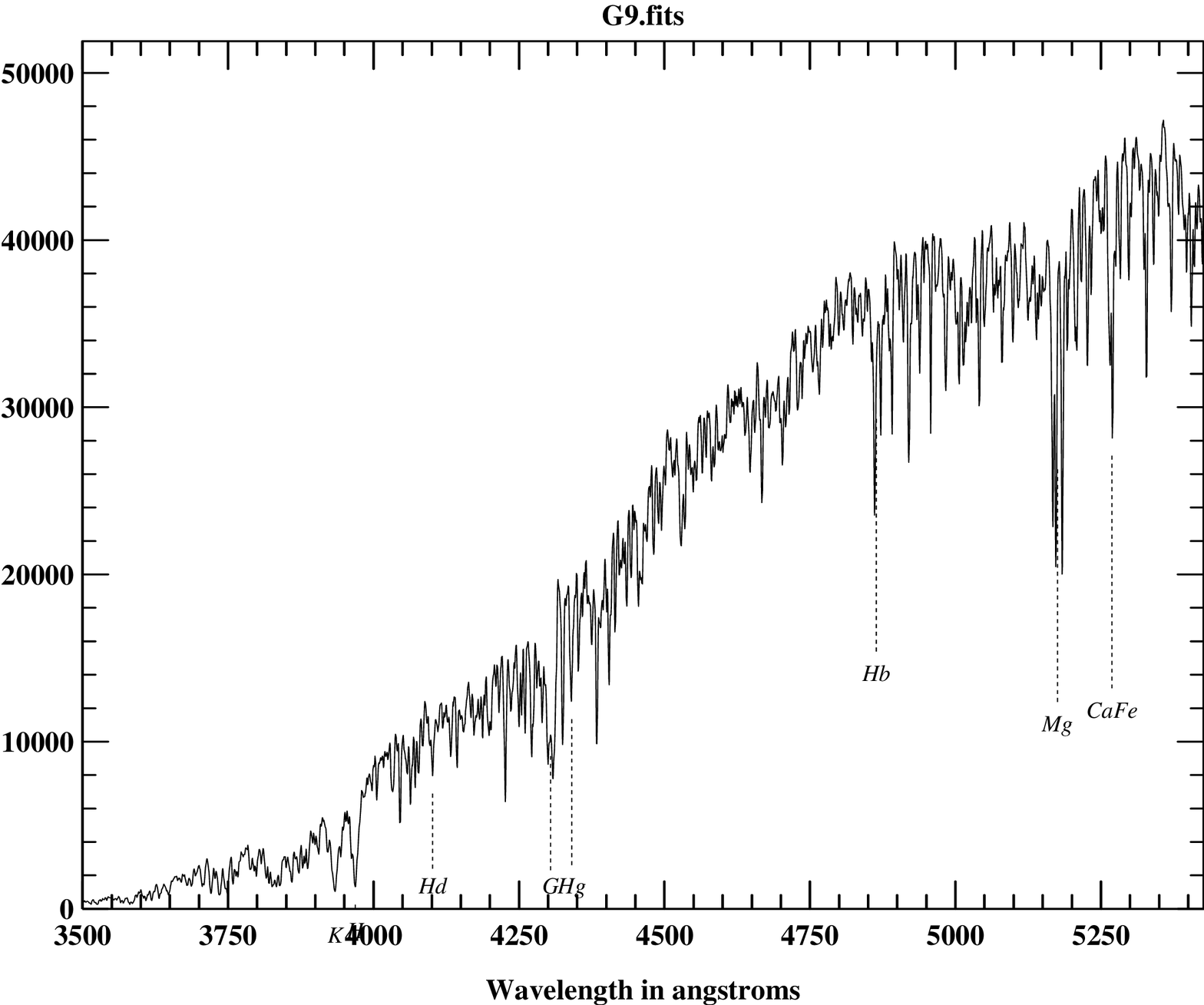}{0.26\textwidth}{e. G9 template.}
          \fig{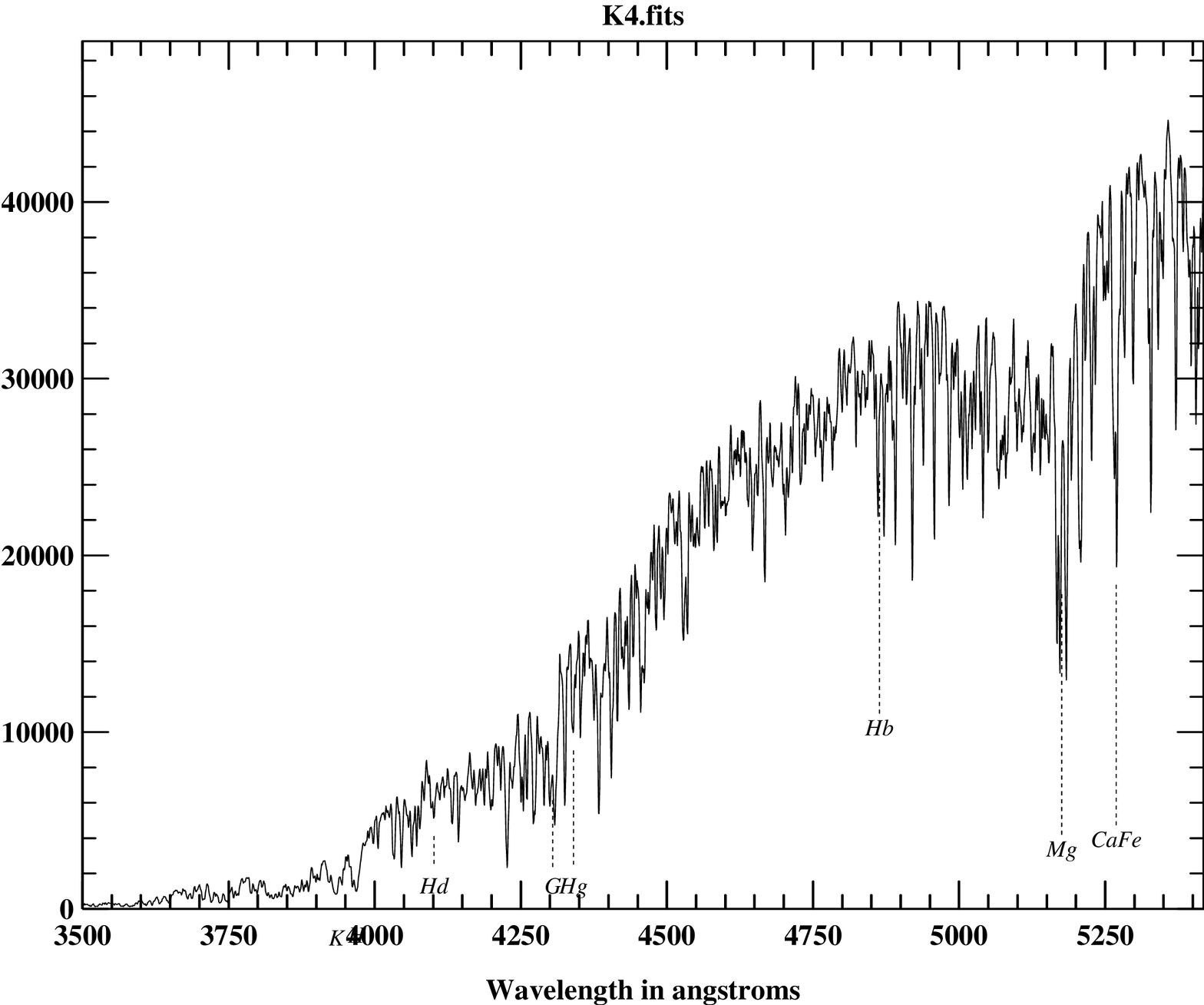}{0.26\textwidth}{f. K4 template.}
          }
\caption{\label{fig:stellartempl}Selected stellar templates for the DISPERSE=600 TILT=445 FAST program.}
\end{figure*}

Figure \ref{fig:commontempl} plots the six most commonly used templates in the FAST archive.  These templates are primarily intended to measure galaxy redshifts from spectra obtained with the standard DISPERSE=300, TILTPOS=600, APERTURE=3 setup.  These six templates account for 75\% of the cross-correlation velocities in the FAST reduced spectra archive.

fglotemp and fs2temp were made in 1994 and 1995, respectively, from FAST globular cluster spectra obtained with the first two CCD detectors.  The K-type spectra were thought to provide a good match for absorption line galaxies.  femtemp97 and fabtemp97 were made in 1997 to better fit galaxy emission and absorption line spectra, respectively.  ztemp was made from absorption line galaxies observed with the old Z-Machine spectrograph.  m31\_a\_temp was made in 2002 from halo A-type stars observed with the MMT Hectospec spectrograph (from a program that targeted M31).  The A-type template provides a better match to flux standard stars and white dwarfs.

Figure \ref{fig:stellartempl} plots selected templates from a series of stellar spectral-type templates, constructed in 2002 from the spectra of bright stars observed with the DISPERSE=600, TILTPOS=445, APERTURE=2.0 setup.  A full series of spectral-type templates was made for PROGRAM=122 \citep{brown03} and are distributed with the RVSAO package; however, the spectra used to make the templates were obtained before 2006 and do not appear in the FAST reduced spectra archive.  The stellar templates in Figure \ref{fig:stellartempl} account for 4\% of the cross-correlation velocities in the FAST reduced spectra archive.

We remind the reader that cross-correlation velocities are accurate only if the template matches the target.  We have no template for certain spectral classes, such as helium stars, M dwarfs, quasars, and supernovae.  Thus, we urge users to analyze FAST spectra with cross-correlation templates optimized for their actual needs.

\begin{deluxetable*}{lrccrccll}
\tablecolumns{9}
\tablecaption{Cross-Correlation Templates\label{tab:fasttemps}}
\tablehead{ \colhead{Template} & \colhead{Nbest} & \colhead{First Night} & \colhead{Last Night} & \colhead{Npix} & \colhead{Range (\AA)} & \colhead{Res.(\AA)} & \colhead{Continuum} & \colhead{Comments} }
\startdata
\cutinhead{Templates for the standard DISPERSE=300 TILT=600 configuration}
eatemp & 1106 & 1994 Jun 05 & 2019 Nov 22 & 4505 & 2979-7000 & 6.0 & Subtracted & Elliptical galaxy+A-star Hectospec \\
eltemp & 947 & 1994 Jun 14 & 2019 Nov 22 & 4505 & 2979-7331 & 6.0 & Instrumental & Elliptical galaxy Hectospec \\
emtemp & 7559 & 1994 Jan 11 & 2019 Dec 22 & 4096 & 3500-7500 & 6.0 & Zero & Galaxy emission lines synthesized \\
f86btemp & 2889 & 1994 Jul 03 & 1995 Jul 31 & 1850 & 3690-6402 & 7.2 & Instrumental & NGC 4486b galaxy 5 FAST \\
fabtemp97 & 26638 & 1994 Jun 05 & 2019 Dec 21 & 2000 & 3800-6700 & 7.2 & Divided & Absorption line galaxy \\
femtemp & 645 & 1994 Jun 05 & 2019 Dec 22 & 20001 & 3000-8000 & 8.0 & Zero & Galaxy emission lines \\
femtemp97 & 39054 & 1994 Jun 05 & 2019 Dec 22 & 10001 & 3000-8000 & 6.0 & Zero & 11 Galaxy emission lines \\
fglotemp & 19920 & 1994 Jan 11 & 2019-12-22 & 4096 & 3700-7500 & 7.2 & Instrumental & M31  cluster 225-280 15 FAST \\
fm32temp & 6420 & 1994 Jan 11 & 2019 Dec 21 & 4096 & 3700-7500 & 7.2 & Instrumental &  M32 cluster stars 14 FAST \\
fn4486btemp & 291 & 1994 Feb 02 & 1996 Jan 23 & 4096 & 3700-7500 & 7.2 & Instrumental & NGC 4486b 23 FAST \\
fn5548temp & 116 & 1994 Feb 02 & 1996 Jan 23 & 4096 & 3700-7300 & 7.2 & Instrumental & Active galaxy NGC 5548 39 FAST \\
fn7331temp & 6723 & 1994 Jan 11 & 2019 Dec 22 & 4096 & 3700-7500 & 7.2 & Instrumental & NGC 7331 galaxy 16 FAST \\
fs2temp & 14486 & 1994 Jan 11 & 2019 Dec 21 & 4096 & 3700-7500 & 7.2 & Instrumental & Star AGK2p4392 53 FAST \\
fztemp97 & 26 & 1997 Jul 30 & 1997 Aug 03 & 2000 & 3800-6700 & 7.2 & Divided & Absorption galaxies 1,979 FAST \\
m31\_a\_temp & 7021 & 1994 Jun 05 & 2019 Dec 22 & 4000 & 3700-7000 & 6.0 & Divided & A stars 39 Hectospec \\
m31\_f\_temp & 1421 & 1994 Jun 09 & 2019 Nov 22 & 4000 & 3700-7000 & 6.0 & Divided & F stars 72 Hectospec \\
m31\_k\_temp & 1362 & 1994 Jun 05 & 2019 Dec 21 & 4000 & 3700-7000 & 6.0 & Divided & K stars 69 Hectospec \\
skytemp & 18 & 1998 Nov 16 & 2014 Apr 30 & 6002 & 4000-7000 & 6.0 & Zero & 10 Night-sky emission lines \\
sptemp & 2010 & 1994 Jun 05 & 2019 Nov 23 & 4505 & 2984-7379 & 6.0 & Instrumental & Spiral galaxy Hectospec \\
stemp & 1231 & 1994 Jul 03 & 1995 Jul 31 & 2048 & 4500-6500 & 4.8 & Divided & Composite star Z-Machine \\
ztemp & 10602 & 1994 Jan 11 & 2019 Dec 22 & 2048 & 4500-6500 & 4.8 & Divided & Composite galaxy Z-machine \\
\cutinhead{Stellar templates for the DISPERSE=600 TILT=445 configuration}
A1 & 840 & 2006 Sep 17 & 2019 Sep 30 & 2679 & 3500-5352 & 2.3 & Instrumental & A1 star combined FAST spectra \\
A4 & 2361 & 2006 Sep 15 & 2019 Sep 30 & 2679 & 3500-5423 & 2.3 & Instrumental & A4 star combined FAST spectra \\
A7 & 241 & 2006 Sep 17 & 2019 Sep 29 & 2683 & 3500-5423 & 2.3 & Instrumental & A7 star combined FAST spectra \\
B3 & 442 & 2002-04-03 & 2019 Sep 29 & 2679 & 3500-5427 & 2.3 & Instrumental & B3 star combined FAST spectra \\
B7 & 615 & 2006 Sep 15 & 2019 Sep 29 & 2679 & 3500-5418 & 2.3 & Instrumental & B7 star combined FAST spectra \\
B9 & 1058 & 2006 Sep 15 & 2019 Sep 30 & 2682 & 3500-5419 & 2.3 & Instrumental & B9 star combined FAST spectra \\
F2 & 99 & 2006 Sep 16 & 2019 Jun 05 & 2679 & 3500-5423 & 2.3 & Instrumental & F2 star combined FAST spectra \\
F4 & 382 & 2008 Apr 15 & 2019 Jul 02 & 2679 & 3500-5420 & 2.3 & Instrumental & F4 star combined FAST spectra \\
F6 & 88 & 2010 Apr 10 & 2019 Mar 14 & 2679 & 3500-5419 & 2.3 & Instrumental & F6 star combined FAST spectra \\
F8 & 44 & 2010 May 08 & 2019 Jul 02 & 2679 & 3500-5423 & 2.3 & Instrumental & F8 star combined FAST spectra \\
F9 & 23 & 2010 May 09 & 2019 Jun 04 & 2679 & 3500-5418 & 2.3 & Instrumental & F9 star combined FAST spectra \\
G2 & 33 & 2010 May 09 & 2019 Jun 05 & 2684 & 3500-5419 & 2.3 & Instrumental & G2 star combined FAST spectra \\
G6 & 17 & 2006 Sep 28 & 2019 Sep 30 & 2679 & 3500-5419 & 2.3 & Instrumental & G6 star combined FAST spectra \\
G9 & 63 & 2010 Sep 09 & 2019 Sep 30 & 2679 & 3500-5426 & 2.3 & Instrumental & G9 star combined FAST spectra \\
K1 & 26 & 2010 Dec 01 & 2019 Sep 29 & 2679 & 3500-5418 & 2.3 & Instrumental & K1 star combined FAST spectra \\
K4 & 80 & 2006 Dec 21 & 2019 Sep 29 & 2679 & 3500-5418 & 2.3 & Instrumental & K4 star combined FAST spectra \\
\enddata
\end{deluxetable*}

\clearpage
\onecolumngrid 
\section{FITS Header Parameters}

The following tables document the keywords in the reduced FITS file headers.  Table \ref{tab:headtel} defines the keywords created by the telescope data acquisition system.  Table \ref{tab:headpipe} defines the keywords added by the data reduction pipeline.  Table \ref{tab:headrv} defines the keywords added by the cross-correlation task, XCSAO, and the confirmation task, QPLOT \citep{tokarz1997, kurtz98}.  The overall listing reflects the order in which the keywords normally appear in the FITS file headers. 

\startlongtable
\begin{deluxetable*}{lcl}
\tablecolumns{3}
\tablewidth{0pt}
\tablecaption{FITS Image Header Keywords Generated by the Telescope Data Acquisition System\label{tab:headtel}}
\tablehead{ \colhead{Keyword} & \colhead{Example Value} & \colhead{Description or Long Example Value} }
\startdata
SIMPLE & T  & Fits standard file \\
BITPIX & -32  & Bits per pixel -32=4-byte floating point \\
NAXIS & 2  & Number of axes  \\
NAXIS1 & 2681  & Dispersion axis length in pixels \\
NAXIS2 & 1 & Number of objects in file \\
NAXIS3 & 4  & Spectra in file: 1=spectrum 2=sky+spectrum 3=sky 4=variance \\
EXTEND & F  & File has no extensions \\
ORIGIN &  & NOAO-IRAF FITS Image Kernel July 2003 \\
DATE & 2013-03-18T19:29:11  & Date this FITS file was generated \\
IRAF-TLM & 2013-03-18T19:42:34  & Time of last modification \\
OBJECT & NGC4116 & Object name \\
NAMPS & 1 & Number of amplifiers read out for original 2-D spectrum \\
DATE-OBS & 2013-03-18  & UT YYYY-MM-DD at start of exposure \\
RA & 12:07:36.81  & Right Ascension \\
DEC & +02:41:32.3  & Declination \\
EPOCH & 2000  & Equinox of coordinates \\
POSANGLE & 90 & Spectrograph position angle [prior to November 12, 2002] \\
ROTANGLE & 90 & Spectrograph position angle [after November 12, 2002] \\
FASTFOC & 1350 & FAST spectrograph focus setting \\
AIRMASS & 1.18 & Airmass \\
RRA & 12:07:42.1  & Requested Right Ascension \\
RDEC & +02:41:32.3 & Requested Declination \\
REPOCH & 2000 & Requested Equinox \\
TRA & 12:07:42.1  & Telescope Right Ascension \\
TDEC & +02:44:03.3  & Telescope Declination \\
TEPOCH & 2000  & Telescope coordinate equinox \\
ST & 11:06:11 & Sidereal time hh:mm:ss at start of exposure \\
HA & -01:02:12 & Hour Angle hh:mm:ss at start of exposure \\
DOME & 117 & Dome azimuth angle in degrees \\
UT & 06:31:30 & UT hh:mm:ss at start of exposure \\
UTEND & 07:01:30 & UT HH:MM:SS at end of exposure \\
ST & '09:27:23'  & Telescope Sidereal Time at end of exposure \\
MJD & 56365.282292 & Modified Julian Date at exposure midtime \\
GJDN & 2456365.782292 & Geocentric Julian Date at exposure midtime \\
HJDN & 2453795.75925  & Heliocentric Julian Date at exposure midtime \\
SITENAME & 'flwo1'  & Observatory name = FLWO 1.5m \\
SITELONG & '+110:52:39.0'  & Observatory longitude in degrees west of 0 ddd:mm:ss \\
SITELAT & '+31:40:51.4'  & Observatory latitude in degrees dd:mm:ss \\
SITEELEV & 2320.0  & Observatory elevation in meters \\
DETECTOR & FAST3 & detector identifier \\
CCDSERIA & STA520A SN4377 &  Chip ID \\
INSTRUME & FAST & instrument name \\
DISKDIR & /data/fast/2013.0313 & Original raw directory with local date yyyy.mmdd \\
DISKFILE & 81 & File number in original raw directory \\
TELESCOP & 'TILLINGHAST'  & 1.5 meter (60 inch) telescope \\
EXPTIME & 1800.000 & Integration time in seconds \\
DARKTIME& 1800.000 & Total elapsed time in seconds \\
IMAGETYP& OBJECT & IRAF spectrum type (object flat comp bias dark) \\
CCDSUM  & 1 2 & Binning of raw image X (dispersion) and Y (cross-dispersion) axes \\
BIN     & 2 & Binning of cross-dispersion axis \\
LTM1\_1  & 1.0000000 & IRAF pixel/pixel in X \\
LTM2\_2  & 1. & IRAF pixel/pixel in Y \\
CCDSEC  & [35:2715,1:161] & Actual detector image limits \\
GAIN    & 0.8000000 & Read gain \\
RDNOISE & 4.4000000 & Read noise \\
OBSERVER& Brown & Telescope observer \\
P.I.    & Zezas & Program principle investigator \\
PROGRAM & & 199 Optical nuclear spectroscopy of nearby galaxies' \\
APERTURE& 3.0 & Slit width on the sky in arcseconds \\
DISPERSE& 300 & Grating lines/mm \\
TILTPOS & 600 & Grating tilt toward blue <600 or red >600 \\
COMMENT &  & Observer: Brown \\
COMMENT &  & P.I.: Zezas \\
COMMENT &  & Program: 199 nearby galaxies nuclear spectroscopy \\
COMMENT &  & Seeing: 1-2" \\
COMMENT &  & Aperture: 3" \\
COMMENT &  & Disperse: 300 \\
COMMENT &  & Tiltpos: 600 \\
COMMENT &  & Focus: 1350 \\
\enddata
\end{deluxetable*}

\pagebreak
\startlongtable
\begin{deluxetable*}{lcl}
\tablecolumns{3}
\tablewidth{0pt}
\tablecaption{FITS Image Header Keywords Generated by the IRAF Data Reduction Pipeline\label{tab:headpipe}}
\tablehead{ \colhead{Keyword} & \colhead{Example Value} & \colhead{Description or Long Example Value} }
\startdata
RFN     & 20130313.0081 & Observation sequence = yyyymmdd.nnnn \\
WCSDIM  & 3 & Number of world coordinate system dimensions \\
WAT0\_001 & system=equispec & IRAF multispec format \\
WAT1\_001 & & wtype=linear label=Wavelength units=angstroms \\
WAT2\_001 & wtype=linear & Cross-dispersion axis \\
TRIM    & & Mar 18 15:12 Trim data section is [35:2715,1:161] \\
OVERSCAN & & Mar 18 15:12 Overscan section is [2:30,1:161] with mean=716.1741 \\
ZEROCOR & & Mar 18 15:12 Zero level correction image is zero \\
CCDMEAN & 155.4683 & Mean value of pixels in CCD image \\
CCDMEANT & 1048086765 & Number of counts in CCD image \\
CCDPROC & & Mar 18 15:12 CCD processing done \\
FLATCOR & & Mar 18 15:12 Flat-field image is norm with scale = 1 \\
DCLOG1  & Transform & Wavelength is transformed from dispersion to linear\\
DC-FLAG & 0 & Wavelength is linear, not log \\
CTYPE1  & LINEAR   & Dispersion axis is linear \\
CTYPE2  & LINEAR   & Cross-dispersion axis is linear \\
CRVAL1  & 3478.38037109375 & Wavelength at center of reference pixel \\
CRPIX1  & 1. & Number of reference pixel in 1D spectrum first is 1 \\
CD1\_1   & 1.47133255004883 & Wavelength per pixel \\
CD2\_2   & 1. & Spectra per pixel \\
BANDID1 & & spectrum - background fit, weights variance, clean yes \\
BANDID2 & & raw - background fit, weights none, clean no \\
BANDID3 & & background - background fit \\
BANDID4 & & sigma - background fit, weights variance, clean yes \\
APNUM1  & 1 1 64.00 119.00 & IRAF APALL extraction info \\
CTYPE3  & LINEAR & Band spectra \\
CD3\_3   & 1. & Band spectra per cross-dispersion pixel \\
LTM3\_3  & 1. & IRAF pixel/pixel in BAND \\
WAT3\_001 & wtype=linear & One band spectrum per cross-dispersion pixel \\ 
FINDOBJ & 85:92 & Cross-dispersion pixel range for object extraction \\
FINDBACK & 65:80,96:111 & Cross-dispersion pixel ranges for background extraction \\
\enddata
\end{deluxetable*}

\pagebreak
\startlongtable
\begin{deluxetable*}{lcl}
\tablecolumns{3}
\tablewidth{0pt}
\tablecaption{FITS Image Header Keywords Generated  by the XCSAO Radial Velocity Program\label{tab:headrv}}
\tablehead{ \colhead{Keyword} & \colhead{Example Value} & \colhead{Description or Long Example Value} }
	\startdata
VELOCITY & 1265.35069460358 & Velocity in km/sec =cz from emission line \\
Z       & 0.00422075500422317 & Velocity as z \\
CZERR   & 300. & Velocity error in km/sec set from emission line\\
ZERR    & 8.83416159802586E-6 & Velocity error in km/sec \\
VELQUAL & Q & Velocity quality Q=good ?=questionable X=bad \\
BCV     & 4.2293119430542 & Solar System barycentric velocity correction \\
CZXC    & 1289.70966527454 &  Cross-correlation velocity = cz\\
ZXC     & 0.00430200777295808 & Cross-correlation z \\
CZXCERR & 2.64841539087617 & Cross-correlation error \\
CZXCR   & 51.4057371762259 & Cross-correlation R-value \\
BESTTEMP & femtemp97.fits & Best cross-correlation template \\
CZPARAMS & 13 4096 1 5 20 250 500 0.05 & No.temps xcor: bins zeropad filterx4 bellwindow \\
COMMENT & & rvsao.xcsao results for each template: \\
COMMENT & & file name, cross-correlation velocity, error, R-value \\
COMMENT & & correlation peak height (0-1) and width (km/sec) \\
COMMENT & & velocity relative to template, template BCV \\
COMMENT & & template em. chopped, template filtered \\
XCTEMP1 & & fabtemp97.fits 1317.375 34.757  6.7242 0.475 350.109 1316.788 0.000 \\
XCTEMP2 & & ztemp.fits 1304.660 19.003 8.48857 0.469 237.782 1300.509 0.000 yes \\
XCTEMP3 & & femtemp97.fits 1289.710 2.648 51.4057 0.877 184.699 1285.354 0.000 n \\
XCTEMP4 & & fs2temp.fits 1281.685 28.899 6.66855 0.463 291.868 1277.337 0.000 ye \\
XCTEMP5 & & fglotemp.fits 1294.446 31.103 6.84486 0.458 319.406 1291.380 0.000 y \\
XCTEMP6 & & fm32temp.fits 1295.454 32.659 6.87543 0.462 336.430 1292.184 0.000 y \\
XCTEMP7 & & fn7331temp.fits 1262.144 48.538 6.40995 0.449 463.554 1257.691 0.000 \\
XCTEMP8 & & eatemp.fits 1328.693 142.354 1.38256 0.098 495.571 1321.587 0.000 no \\
XCTEMP9 & & eltemp.fits 99085.439 292.865 1.71597 0.118 1093.522 99079.649 0.000 \\
XCTEMP10 & & sptemp.fits 1304.153 6.535 23.2321 0.819 210.281 1271.864 -28.010 no \\
XCTEMP11 & & m31\_a\_temp.fits 100184.771 379.495 1.10537 0.080 1304.335 37205.074  \\
XCTEMP12 & & m31\_f\_temp.fits 99610.749 148.512 1.59509 0.100 497.927 15643.079 0. \\
XCTEMP13 & & m31\_k\_temp.fits 77365.998 123.226 1.73377 0.124 618.769 86365.812 0. \\
XCSAO   & & rvsao.xcsao 2.8.2  18-Mar-2013 15:34 CZXC = 1289.71 R = 51.41 \\
VELSET  & & Velocity from 6730.815A S2 emission line \\
QPLOT   & & rvsao.xcsao 2.8.2  18-Mar-2013 15:43 VELOCITY = 1265.35 Q = Q \\
\enddata
\end{deluxetable*}

\clearpage

\bibliography{fast}

\begin{thebibliography}{}
\expandafter\ifx\csname natexlab\endcsname\relax\def\natexlab#1{#1}\fi
\providecommand{\url}[1]{\href{#1}{#1}}

\bibitem[{{Blondin} {et~al.}(2012){Blondin}, {Matheson}, {Kirshner}, {Mand el},
  {Berlind}, {Calkins}, {Challis}, {Garnavich}, {Jha}, {Modjaz}, {Riess}, \&
  {Schmidt}}]{blondin12}
{Blondin}, S., {Matheson}, T., {Kirshner}, R.~P., {et~al.} 2012, \aj, 143, 126

\bibitem[{{Brown} {et~al.}(2003){Brown}, {Allende Prieto}, {Beers}, {Wilhelm},
  {Geller}, {Kenyon}, \& {Kurtz}}]{brown03}
{Brown}, W.~R., {Allende Prieto}, C., {Beers}, T.~C., {et~al.} 2003, \aj, 126,
  1362

\bibitem[{{Brown} {et~al.}(2008){Brown}, {Beers}, {Wilhelm}, {Allende Prieto},
  {Geller}, {Kenyon}, \& {Kurtz}}]{brown08}
{Brown}, W.~R., {Beers}, T.~C., {Wilhelm}, R., {et~al.} 2008, \aj, 135, 564

\bibitem[{{Brown} {et~al.}(2010){Brown}, {Kilic}, {Allende Prieto}, \&
  {Kenyon}}]{brown10}
{Brown}, W.~R., {Kilic}, M., {Allende Prieto}, C., \& {Kenyon}, S.~J. 2010,
  \apj, 723, 1072

\bibitem[{{Carter} {et~al.}(2001){Carter}, {Fabricant}, {Geller}, {Kurtz}, \&
  {McLean}}]{carter01}
{Carter}, B.~J., {Fabricant}, D.~G., {Geller}, M.~J., {Kurtz}, M.~J., \&
  {McLean}, B. 2001, \apj, 559, 606

\bibitem[{{Chilingarian} {et~al.}(2004){Chilingarian}, {Bartunov}, {Richter},
  \& {Sigaev}}]{2004ASPC..314..225C}
{Chilingarian}, I., {Bartunov}, O., {Richter}, J., \& {Sigaev}, T. 2004, in
  Astronomical Society of the Pacific Conference Series, Vol. 314, Astronomical
  Data Analysis Software and Systems (ADASS) XIII, ed. F.~{Ochsenbein}, M.~G.
  {Allen}, \& D.~{Egret}, 225

\bibitem[{{C{\'u}neo} {et~al.}(2018){C{\'u}neo}, {Kenyon}, {G{\'o}mez},
  {Chochol}, {Shugarov}, \& {Kolotilov}}]{cuneo18}
{C{\'u}neo}, V.~A., {Kenyon}, S.~J., {G{\'o}mez}, M.~N., {et~al.} 2018, \mnras,
  479, 2728

\bibitem[{{Demleitner} {et~al.}(2014){Demleitner}, {Neves}, {Rothmaier}, \&
  {Wambsganss}}]{2014A&C.....7...27D}
{Demleitner}, M., {Neves}, M.~C., {Rothmaier}, F., \& {Wambsganss}, J. 2014,
  Astronomy and Computing, 7, 27

\bibitem[{{Dower} {et~al.}(2018){Dower}, {Demleitner}, {Benson}, {Plante},
  {Auden}, {Graham}, {Greene}, {Hill}, {Linde}, {Morris}, {O`Mullane}, {Rixon},
  {St{\'e}b{\'e}}, \& {Andrews}}]{2018ivoa.spec.0723D}
{Dower}, T., {Demleitner}, M., {Benson}, K., {et~al.} 2018, Registry Interfaces
  Version 1.1, IVOA Recommendation 23 July 2018, , ,
  doi:10.5479/ADS/bib/2018ivoa.spec.0723D

\bibitem[{{Fabricant} {et~al.}(1998){Fabricant}, {Cheimets}, {Caldwell}, \&
  {Geary}}]{fabricant1998}
{Fabricant}, D., {Cheimets}, P., {Caldwell}, N., \& {Geary}, J. 1998, \pasp,
  110, 79

\bibitem[{{Fabricant} {et~al.}(2019){Fabricant}, {Fata}, {Epps}, {Gauron},
  {Mueller}, {Zajac}, {Amato}, {Barberis}, {Bergner}, {Brennan}, {Brown},
  {Chilingarian}, {Geary}, {Kradinov}, {McLeod}, {Smith}, \&
  {Woods}}]{2019PASP..131g5004F}
{Fabricant}, D., {Fata}, R., {Epps}, H., {et~al.} 2019, \pasp, 131, 075004

\bibitem[{{Falco} {et~al.}(1999){Falco}, {Kurtz}, {Geller}, {Huchra}, {Peters},
  {Berlind}, {Mink}, {Tokarz}, \& {Elwell}}]{falco99}
{Falco}, E.~E., {Kurtz}, M.~J., {Geller}, M.~J., {et~al.} 1999, \pasp, 111, 438

\bibitem[{{Geller} \& {Huchra}(1989)}]{geller89}
{Geller}, M.~J., \& {Huchra}, J.~P. 1989, Science, 246, 897

\bibitem[{{Hern{\'a}ndez} {et~al.}(2005){Hern{\'a}ndez}, {Calvet}, {Hartmann},
  {Brice{\~n}o}, {Sicilia-Aguilar}, \& {Berlind}}]{hernandez05}
{Hern{\'a}ndez}, J., {Calvet}, N., {Hartmann}, L., {et~al.} 2005, \aj, 129, 856

\bibitem[{{Hicken} {et~al.}(2017){Hicken}, {Friedman}, {Blondin}, {Challis},
  {Berlind}, {Calkins}, {Esquerdo}, {Matheson}, {Modjaz}, {Rest}, \&
  {Kirshner}}]{hicken17}
{Hicken}, M., {Friedman}, A.~S., {Blondin}, S., {et~al.} 2017, \apjs, 233, 6

\bibitem[{{Huchra} {et~al.}(2012){Huchra}, {Macri}, {Masters}, {Jarrett},
  {Berlind}, {Calkins}, {Crook}, {Cutri}, {Erdo{\v g}du}, {Falco}, {George},
  {Hutcheson}, {Lahav}, {Mader}, {Mink}, {Martimbeau}, {Schneider},
  {Skrutskie}, {Tokarz}, \& {Westover}}]{huchra12}
{Huchra}, J.~P., {Macri}, L.~M., {Masters}, K.~L., {et~al.} 2012, \apjs, 199,
  26

\bibitem[{{Kenyon} \& {Garcia}(2016)}]{kenyon2016}
{Kenyon}, S.~J., \& {Garcia}, M.~R. 2016, \aj, 152, 1

\bibitem[{{Kurtz} \& {Mink}(1998)}]{kurtz98}
{Kurtz}, M.~J., \& {Mink}, D.~J. 1998, \pasp, 110, 934

\bibitem[{{Kurtz} \& {Mink}(2000)}]{kurtz2000}
---. 2000, \apjl, 533, L183

\bibitem[{{Kurtz} \& {Mink}(1999)}]{kurtz1999}
{Kurtz}, M.~J., \& {Mink}, J. 1999, {RVSAO 2.0: Digital Redshifts and Radial
  Velocities}, Astrophysics Source Code Library, , , ascl:9912.003

\bibitem[{{Lazo} {et~al.}(2018){Lazo}, {Zahid}, {Sohn}, \& {Geller}}]{lazo18}
{Lazo}, B., {Zahid}, H.~J., {Sohn}, J., \& {Geller}, M.~J. 2018, rnaas, 2, 234

\bibitem[{{Mahdavi} \& {Geller}(2004)}]{mahdavi04}
{Mahdavi}, A., \& {Geller}, M.~J. 2004, \apj, 607, 202

\bibitem[{{Massey} {et~al.}(1988){Massey}, {Strobel}, {Barnes}, \&
  {Anderson}}]{massey88}
{Massey}, P., {Strobel}, K., {Barnes}, J.~V., \& {Anderson}, E. 1988, \apj,
  328, 315

\bibitem[{{Matheson} {et~al.}(2008){Matheson}, {Kirshner}, {Challis}, {Jha},
  {Garnavich}, {Berlind}, {Calkins}, {Blondin}, {Balog}, {Bragg}, {Caldwell},
  {Dendy Concannon}, {Falco}, {Graves}, {Huchra}, {Kuraszkiewicz}, {Mader},
  {Mahdavi}, {Phelps}, {Rines}, {Song}, \& {Wilkes}}]{matheson08}
{Matheson}, T., {Kirshner}, R.~P., {Challis}, P., {et~al.} 2008, \aj, 135, 1598

\bibitem[{{McLeod} {et~al.}(2012){McLeod}, {Fabricant}, {Nystrom}, {McCracken},
  {Amato}, {Bergner}, {Brown}, {Burke}, {Chilingarian}, {Conroy}, {Curley},
  {Furesz}, {Geary}, {Hertz}, {Holwell}, {Matthews}, {Norton}, {Park}, {Roll},
  {Zajac}, {Epps}, \& {Martini}}]{2012PASP..124.1318M}
{McLeod}, B., {Fabricant}, D., {Nystrom}, G., {et~al.} 2012, \pasp, 124, 1318

\bibitem[{{Mink}(2002)}]{mink2002}
{Mink}, D.~J. 2002, in Astronomical Data Analysis Software and Systems XI, Vol.
  281, 169

\bibitem[{{Mink}(2013)}]{mink2013}
{Mink}, J. 2013, in Astronomical Society of the Pacific Conference Series, Vol.
  475, Astronomical Data Analysis Software and Systems XXII, ed. D.~N.
  {Friedel}, 291

\bibitem[{{Mink} {et~al.}(2019){Mink}, {Rhee}, \& {Latham}}]{mink2016}
{Mink}, J., {Rhee}, J., \& {Latham}, D.~W. 2019, in Astronomical Society of the
  Pacific Conference Series, Vol. 521, Astronomical Data Analysis Software and
  Systems XXVI, ed. M.~{Molinaro}, K.~{Shortridge}, \& F.~{Pasian}, 132

\bibitem[{{Raddi} {et~al.}(2015){Raddi}, {Drew}, {Steeghs}, {Wright}, {Drake},
  {Barentsen}, {Fabregat}, \& {Sale}}]{raddi15}
{Raddi}, R., {Drew}, J.~E., {Steeghs}, D., {et~al.} 2015, \mnras, 446, 274

\bibitem[{{Rines} {et~al.}(2003){Rines}, {Geller}, {Kurtz}, \&
  {Diaferio}}]{rines03}
{Rines}, K., {Geller}, M.~J., {Kurtz}, M.~J., \& {Diaferio}, A. 2003, \aj, 126,
  2152

\bibitem[{{Taylor}(2005)}]{2005ASPC..347...29T}
{Taylor}, M.~B. 2005, in Astronomical Society of the Pacific Conference Series,
  Vol. 347, Astronomical Data Analysis Software and Systems XIV, ed.
  P.~{Shopbell}, M.~{Britton}, \& R.~{Ebert}, 29

\bibitem[{{Tody} {et~al.}(2011){Tody}, {Micol}, {Durand }, {Louys}, {Bonnarel},
  {Schade}, {Dowler}, {Michel}, {Salgado}, {Chilingarian}, {Rino}, {de Dios
  Santand er}, \& {Skoda}}]{2011ivoa.spec.1028T}
{Tody}, D., {Micol}, A., {Durand }, D., {et~al.} 2011, Observation Data Model
  Core Components, its Implementation in the Table Access Protocol Version 1.0,
  IVOA Recommendation 28 October 2011, , ,
  doi:10.5479/ADS/bib/2011ivoa.spec.1028T

\bibitem[{{Tody} {et~al.}(2012){Tody}, {Dolensky}, {McDowell}, {Bonnarel},
  {Budavari}, {Busko}, {Micol}, {Osuna}, {Salgado}, {Skoda}, {Thompson},
  {Valdes}, \& {Data Access Layer Working Group}}]{2012ivoa.spec.0210T}
{Tody}, D., {Dolensky}, M., {McDowell}, J., {et~al.} 2012, Simple Spectral
  Access Protocol Version 1.1, IVOA Recommendation 10 February 2012, , ,
  doi:10.5479/ADS/bib/2012ivoa.spec.0210T

\bibitem[{{Tokarz} \& {Roll}(1997)}]{tokarz1997}
{Tokarz}, S.~P., \& {Roll}, J. 1997, in Astronomical Society of the Pacific
  Conference Series, Vol. 125, Astronomical Data Analysis Software and Systems
  VI, ed. G.~{Hunt} \& H.~{Payne}, 140

\bibitem[{{Tonry} \& {Davis}(1979)}]{tonry1979}
{Tonry}, J., \& {Davis}, M. 1979, \aj, 84, 1511

\bibitem[{{Trichas} {et~al.}(2012){Trichas}, {Green}, {Silverman}, {Aldcroft},
  {Barkhouse}, {Cameron}, {Constantin}, {Ellison}, {Foltz}, {Haggard},
  {Jannuzi}, {Kim}, {Marshall}, {Mossman}, {P{\'e}rez}, {Romero-Colmenero},
  {Ruiz}, {Smith}, {Smith}, {Torres}, {Wik}, {Wilkes}, \&
  {Wolfgang}}]{trichas12}
{Trichas}, M., {Green}, P.~J., {Silverman}, J.~D., {et~al.} 2012, \apjs, 200,
  17

\end{thebibliography}

\end{document}